\documentclass[openacc]{rstransa}

\usepackage{bm, url}
\usepackage{comment}
\usepackage{ifthen}
\usepackage{subfigure}
\usepackage{soul}
\usepackage[normalem]{ulem}

\newboolean{showcomments}
\setboolean{showcomments}{true}

\newcommand{\dennice}[1]{  \ifthenelse{\boolean{showcomments}}
{\textcolor{purple}{(Dennice says:  #1)}}{}}
\newcommand{\brian}[1]{  \ifthenelse{\boolean{showcomments}}
{\textcolor{red}{(Brian says:  #1)}}{}}
\newcommand{\petros}[1]{  \ifthenelse{\boolean{showcomments}}
{\textcolor{red}{(Petros says:  #1)}}{}}

\newcommand{\addcites}[0]{\ifthenelse{\boolean{showcomments}}
{\textcolor{blue}{(add cite(s))}}{}}

\definecolor{mypink1}{rgb}{0.858, 0.188, 0.478}
\newcommand{\upi}{\pi}

\newcommand\Cm{{\bf{C}}}

\newcommand\Qm{{\bf{Q}}}

\definecolor{ForestGreen}{RGB}{34,139,34}
\definecolor{purple}{RGB}{160,32,240}

\def\uv{\mathbf{u}}
\def\u{\mathbf{u}}

\def\xv{\mathbf{x}} 
\def\fv{\mathbf{f}} 

\def\A{\bm{\mathsf{A}}}

\def\A{\bm{\mathsf{A}}}

\newcommand{\be}{\begin{equation}}
\newcommand{\ee}{\end{equation}}
\newcommand{\bdm}{\begin{equation*}}
\newcommand{\edm}{\end{equation*}}
\newcommand{\bea}{\begin{eqnarray}}
\newcommand{\eea}{\end{eqnarray}}

\newcommand{\partialf}[2]
{
 \ifthenelse{\equal{#1}{}}{\frac{\partial}{\partial #2}}{\frac{\partial #1}{\partial #2}}
}

\newcommand{\df}{\textrm{d}}

\providecommand\bcdot{\boldsymbol{\cdot}}

\newcounter{saveeqn}%
\begin{document}

\title{A statistical state dynamics approach to wall-turbulence}

\author{
B. F. Farrell$^{1}$, D. F. Gayme$^2$, and P. J. Ioannou$^{3}$}

\address{$^{1}$Harvard University\\
$^{2}$Johns Hopkins University\\
$^{3}$National Kapodistrian University of Athens
}

\subject{xxxxx, xxxxx, xxxx}


\corres{Brian F. Farrell\\
\email{farrell@seas.harvard.edu}}

\begin{abstract}

{\color{black} This paper reviews  results  from the study of  wall-bounded turbulent flows using  statistical state dynamics  (SSD)
that demonstrate  the benefits of adopting this  perspective for understanding  turbulence in wall-bounded shear flows}.
The SSD approach used in this work employs a second-order closure which isolates the interaction between the
streamwise mean and the equivalent of the perturbation covariance.
This closure restricts nonlinearity in the SSD
to that explicitly retained in the streamwise constant mean  together with nonlinear interactions between the mean  and
the perturbation covariance.
This dynamical restriction, in which explicit perturbation--perturbation nonlinearity is removed from the pertur-bation
equation, results in a simplified dynamics referred to as the
restricted nonlinear (RNL) dynamics.
RNL systems in which  an ensemble of a finite number of  realizations of the perturbation equation share the same mean flow provide  tractable approximations to  the equivalently infinite ensemble
RNL system. The infinite ensemble system, referred to as the S3T, introduces new analysis tools for studying turbulence.  The  RNL with a single ensemble member
can be alternatively viewed as a realization of RNL dynamics.
RNL systems provide computationally efficient means to approximate the SSD, producing self-sustaining turbulence
exhibiting qualitative fea-tures similar to those observed in direct numerical simulations (DNS) despite its greatly simplified dynamics.
Finally, we show that RNL turbulence can be supported by as few as a single streamwise varying component interacting with the streamwise
constant mean flow and that judicious selection of this truncated support, or `band-limiting', can be used to improve quantitative accuracy of RNL turbulence.
The results suggest that the SSD approach provides new analytical and computational tools allowing new insights into wall-turbulence.

\end{abstract}

\maketitle

\section{Introduction}
\label{sec:intro}

Wall-turbulence plays a critical role in a wide range of engineering and physics problems.  Despite the
acknowledged importance of improving understanding of wall-turbulence and an extensive literature recording advances in the study of this problem,  fundamental aspects of wall-turbulence remain unresolved. The enduring challenge of understanding  turbulence can be partially attributed to the fact that the Navier-Stokes (NS) equations, which are known to govern its dynamics, are analytically intractable.  Even though there has been a great deal of progress in simulating turbulence, see e.g. Refs. \cite{Kim-etal-1987,DelAlamo-etal-2004,Tsukahara-etal-2006, Wu-Moin-2009,Scott-Polvani-2008,Lee-Moser-2015}, a complete understanding of the physical mechanisms underlying turbulence remains elusive.
This challenge has motivated the search for analytically simpler and computationally
more tractable dynamical models that retain the fundamental mechanisms of turbulence while
facilitating insight into the underlying dynamics and providing a simplified platform for
computation.  A  statistical state dynamics (SSD)  model
comprising coupled evolution equations for a mean flow and a perturbation covariance provides a
new framework for analyzing the dynamics of turbulence.
The {\color{black} restricted nonlinear (RNL)} approximation in which the perturbation covariance is replaced by a finite number of realizations of the perturbation equation that share the same mean flow provide complementary tools for tractable computations.


The use of statistical variables is well accepted as an approach to analyzing complex spatially and temporally varying fields arising in physical systems and  analyzing observations and simulations of turbulent systems using statistical quantities is common practice.  However, it is less common to adopt statistical  variables explicitly for expressing the dynamics of the turbulent system.  An early attempt to exploit the potential of employing statistical state dynamics
directly to provide insight into the mechanisms underlying turbulence involved formal expansion of the equations in cumulants \cite{Hopf-1952,Frisch-1995}. Despite its being an important conceptual advance, the cumulant method was subsequently restricted in application, in  part due to the difficulty of obtaining robust closure of the expansion when it was applied to isotropic homogeneous turbulence.  Another familiar example of a theoretical application of statistical state dynamics (SSD) to turbulence is provided by the Fokker-Planck equation. Although this expression of SSD is insightful, attempting to use it to evolve high dimensional dynamical systems
leads to intractable  representations of the associated SSD.
These examples illustrate one of the key reasons SSD methods have remained underexploited; the assumption that obtaining the dynamics of the statistical states is prohibitively difficult in practice. This perceived difficulty of implementing SSD to study systems of the type
typified by turbulent flows, has led to a focus on realizations of  state trajectories and then analyzing the results to obtain an approximation to the assumed statistically steady
probability density function of the turbulent state or to compile approximations to the statistics of variables. However, this emphasis on  realizations
of the  dynamics has at least one critical limitation: it fails to provide
 insight into phenomena that are intrinsically associated  with the dynamics of the statistical state,  which is a concept distinct from the dynamics of
individual realizations. While the role of multiscale cooperative phenomena involved in the dynamics of
turbulence is often compellingly apparent in the statistics  of realizations, the cooperative phenomena involved  influence
the trajectory  of the statistical state of the system, the evolution of which is controlled by  its statistical state dynamics.
{\color{black} For example, stability analysis of the
statistical state dynamics   associated with  barotropic and baroclinic beta-plane turbulence predicts spontaneous formation of jets with the observed structure.
These results are consistent with
jet formation  and maintenance observed  in the atmospheres of the gaseous planets  arising from  an unstable mode of the statistical state dynamics
that has no analytical  counterpart in realization dynamics.  This jet formation instability has clear connections to observed behavior, so while jet formation is clear in realizations it cannot be comprehensively understood within the framework of  realization dynamics \cite{Farrell-Ioannou-2003-structural,Farrell-Ioannou-2007-structure, Farrell-Ioannou-2009-closure, Bakas-Ioannou-2011,Bakas-Ioannou-2013-prl,Srinivasan-Young-2012}. This example demonstrates how
statistical state dynamics can bring conceptual clarity to the study of turbulence.   This clarification of concept and associated deepening of understanding of turbulence dynamics constitutes an important contribution of the statistical state dynamics perspective.  }

In this work we focus on the study of wall turbulence.
The mean flow is  taken to be the streamwise averaged flow \cite{Gayme-etal-2010}, and the perturbations are the deviations from this mean.  Restriction of the dynamics to the first two cumulants  involves either parameterizing the third cumulant by  stochastic excitation \cite{Farrell-Ioannou-1993e,DelSole-Farrell-1996,DelSole-04} or, as we will adopt in this work,  setting it to zero \cite{Thomas2014, Marston-etal-2008,Tobias-etal-2011,Srinivasan-Young-2012}.  Either of these closures results in retaining only interaction between the perturbations and the mean while neglecting explicit calculation of the perturbation-perturbation interactions.
 This closure results in nonlinear  evolution equations for  the statistical state of the turbulence comprising the mean flow and the second-order perturbation statistics.  If the system being studied has sufficiently low dimension these second-order perturbation statistics can be obtained from a time dependent matrix Lyapunov equation corresponding to an infinite ensemble of realizations.   Results obtained from studying jet formation in 2D planetary dynamics and more recent results in which statistical state dynamics methods were applied to study low Reynolds number
wall-turbulence \cite{Farrell-Ioannou-2012}  motivated further work in analyzing and
simulating turbulence by directly exploiting  statistical state dynamics methods and concepts as an alternative to the traditional approach of studying the dynamics of single realizations.   However, an impediment to the project of extending statistical state dynamics methods to higher Reynolds number turbulence soon became
apparent: because  the second cumulant is of dimension $N^2$ for a system of dimension $N$
direct integration of the statistical state dynamics equations is limited to relatively low resolution systems and therefore low
Reynolds numbers.  In this report  the focus is on methods for extending application of the statistical state dynamics (SSD)
approach by exploiting the restricted nonlinear (RNL) model, which has  recently shown success in the study of a wide range of flows, see eg. Refs. \cite{Farrell-Ioannou-2012,Constantinou-etal-2014b, Thomas2014,Thomas2015, Bretheim-etal-2015, Farrell-etal-2016-VLSM}. The RNL model implementations of SSD comprise joint evolution of a coherent mean flow (first cumulant) and an ensemble approximation to the second-order perturbation statistics which is considered conceptually to be an approximation to the covariance of the perturbations (second  cumulant) although this covariance is not explicitly calculated.

One reason the statistical state dynamics modeling framework provides an appealing tool for studying the maintenance and regulation of turbulence is that RNL turbulence naturally gives rise to a ``minimal realization'' of the dynamics \cite{Farrell-Ioannou-2012, Thomas2015}.  This ``minimal realization'' does not rely on a particular  Reynolds number or result from restricting the channel size and therefore Reynolds number trends as well as the effects of increasing the channel size can be explored within the RNL framework.  A second advantage of the RNL framework is that it does not model particular flow features, such as the roll and the streak, in isolation but rather captures the dynamics of these structures as part of the holistic turbulent dynamics \cite{Thomas2014}.

\section{A statistical state dynamics model for wall-bounded turbulence}
\label{sec:framework}

Consider a {\color{black} parallel wall-bounded shear flow} with streamwise direction $x$, wall-normal direction  $y$, and  spanwise direction $z$ with respective channel  extents in the streamwise, wall-normal, and spanwise  direction $L_x$, $2\delta$ and $L_z$.
 The non-dimensional Navier-Stokes equations (NS)  governing the dynamics assuming a  uniform unit
density  incompressible fluid are:

\begin{equation}
\partial_t \uv^{tot}+   \uv^{tot} \cdot \nabla \uv^{tot} =- \nabla {p} +  \Delta { \uv^{tot}} / Re - (\partial_x  p_\infty ) \hat{\xv} +\fv  ~,{\rm with} ~~\nabla \cdot \uv^{tot} = 0~,
\label{eq:NSE0}
\end{equation}
where  $\uv^{tot} (\xv , t)$ is the velocity field, $p(\xv ,t)$ {\color{black} is the  pressure field,
$\hat{\xv}$ is }the unit vector in the $x$ direction,  {\color{black}  and $\fv$ is a divergence-free
external excitation.
In the non-dimensional eq.  \eqref{eq:NSE0} velocities have been scaled by the characteristic velocity of the laminar flow $U_m$, lengths by the characteristic length $\delta$,  and time by $\delta/U_m$, and  $Re=U_m \delta/\nu$
is the Reynolds number with kinematic viscosity $\nu$. The velocity scale $U_m$ is  specified according to the flow configuration of interest.  For example, \eqref{eq:NSE0} with no imposed pressure gradient $(i.e.,(\partial_x  p_\infty ) \hat{\xv}=0)$, $U_m$ equal to half the maximum velocity difference across a channel with walls at $y/\delta= \pm 1$ and  boundary conditions $\u^{tot}(x, \pm 1, z) = \pm \hat \xv$ describes a plane Couette flow.  Equation \eqref{eq:NSE0} with a constant pressure gradient $\partial_x p_{\infty} $, a characteristic velocity scale $U_m$ equal to the centerline or bulk velocity (for the laminar flow) with boundary conditions $\u^{tot}(x,\pm 1,z)=0$ describes a  Poiseuille (channel) flow.  Throughout this work, we impose periodic
boundary conditions in the streamwise and spanwise directions. We discuss how this equation can be used to represent a half-channel flow  in Section \ref{sec:RNL_halfchannel}. }

%
%

Pressure can be eliminated from these equations and nondivergence enforced through the use of the Leray projection operator,
$P_L(\cdot)$
\cite{Foias-etal-2001}.
Using the Leray projection  the NS expressed in velocity variables become\footnote{The Leray projection annihilates the gradient of a scalar field. For this reason the  $p_\infty$  term does not appear in the  projected equations.}  :
\begin{equation}
\label{eq:NSEPL}
{\partial_t \u^{tot}}+  P_L \left ( \u^{tot} \cdot \nabla \u^{tot}  -  \Delta { \u^{tot}}/{ Re}  \right )= \fv ~.
\end{equation}

%

Obtaining equations for the statistical state dynamics of channel flow requires  an averaging operator, denoted
with angle brackets, $\langle \;\bcdot\;\rangle$, which satisfies  the Reynolds conditions:
\be
\label{eq:rey}
\langle  \alpha f  + \beta g \rangle = \alpha \langle f \rangle + \beta \langle g \rangle,  ~\langle \partial_t f \rangle =
\partial_t \langle f\rangle, ~~\langle \langle f \rangle g \rangle = \langle f \rangle \langle g \rangle ~,
\ee
in which $f(\xv,t)$ and $g(\xv,t)$ are flow variables  and   $\alpha$,$\beta$ are constants (cf. section 3.1 \cite{Monin-Yaglom-1971}).
The  statistical state dynamics variables  are the spatial cumulants of the velocity.
In contrast to the statistical state dynamics of isotropic and homogeneous turbulence, the statistical state dynamics of
 wall-bounded turbulence can be well approximated by retaining only the first two cumulants \cite{Farrell-Ioannou-2012}.
 The first cumulant of the flow field is the mean velocity, ${\bf U}\equiv\langle \u^{tot} \rangle$,  with components $(U,V,W)$, while the second is the  covariance of the perturbation velocity,  $\u=\u^{tot} - {\bf U}$,
 between two spatial points, $\mathbf{x}_1$ and $\mathbf{x}_2$:  $C_{ij} (1,2)\equiv\langle u_i (\xv_1, t) u_j(\xv_2,t) \rangle$.

Averaging operators  satisfying the Reynolds conditions   include ensemble averages and spatial
 averages over coordinates.   Spatial averages will be denoted  by  angle brackets with a subscript indicating  the independent variable over which the average is taken, i.e.~streamwise  averages by $ \langle\,\boldsymbol{\cdot}\,\rangle_x=L_x^{-1} \int_0^{L_x} \boldsymbol{\cdot}\ \df x$
 and averages in both the streamwise and spanwise  by $\langle\,\boldsymbol{\cdot}\,\rangle_{x,z}$.
 Temporal averages will be indicated by an overline, $\overline{\;\boldsymbol{\cdot}\;} = \tfrac{1}{T} \int_0^T \boldsymbol{\cdot}~{\rm{d}} t$,
 with  $T$ sufficiently large.

An important consideration in the study of turbulence  using statistical state dynamics  is choosing an averaging operator that isolates the primary coherent motions.  The associated closure must also maintain the interactions between the coherent mean and incoherent perturbation structures that determine the physical mechanisms underlying the turbulence dynamics.   \iffalse that isolate the dynamically primary mean structures while permitting the physical mechanism of interaction between the coherent mean and incoherent perturbation structures.}  \dennice{The previous sentence is rather confusing an needs to be rewritten.  the averaging operator does not really affect the interactions directly (that comes through the joint evolution of the mean and the covariance) and the following text explains the main ideas}\fi
The detailed structure of the coherent components is critical  in producing
energy transfer from the externally forced flow to the perturbations, therefore
retaining the nonlinearity and structure of the  mean flow components is crucial.  In contrast,  nonlinearity and comprehensive structure  information is  not required to account for the role of the
incoherent motions so that the statistical information contained in the second cumulant suffices to
include the influence of the perturbations on the turbulence  dynamics. Retaining the complete structure and  dynamics of the coherent component while retaining only the necessary statistical correlation for the incoherent component  results in a great practical as well as conceptual simplification.

In the case of  wall-bounded shear flow
there is a great deal of experimental and analytical evidence indicating the prevalence and central role of streamwise elongated coherent structures, see e.g. Refs.
\cite{Kim-Adrian-1999,Guala-etal-2006,Hutchins_Marusic-2007,Kline-et-al-1967,Komminaho-etal-1996,Blackwelder-Kaplan-1976,Bullock-etal-1978, Farrell-Ioannou-1993e,Jovanovic-Bamieh-2005,Cossu-etal-2009,Hwang-Cossu-2010}.
It is of particular importance that the mean flow dynamics capture the interactions between streamwise elongated streak and roll structures in the
self-sustaining process (SSP) \cite{Waleffe-1997,Jimenez-Pinelli-1999,Hamilton-etal-1995,Hwang-Cossu-2010a,Hwang-Cossu-2010b}.
 Streamwise constant models\cite{ Reynolds-Kassinos-1995,Bobba-etal-2002,BobbaThesis}, which implicitly simulate these structures  have been shown to capture  components of mechanisms such as the
 nonlinear momentum transfer and associated increased wall shear stress characteristic
of wall-turbulence~\cite{Gayme-etal-2010, GaymeThesis,Gayme-etal-2011,Bourguignon-McKeon-2011}.
On the other hand, taking the mean over both homogeneous directions ($x$ and $z$) does not capture the roll/streak SSP dynamics  and  this mean does not result   in a second-order closure that maintains
turbulence ~\cite{Jimenez-Pinelli-1999}.

We therefore select $\mathbf{U}=\langle \u^{tot} \rangle_x$ as the first cumulant, which leads to a streamwise
constant mean flow which captures the dynamics of coherent roll/streak  structures.
We define the streak component of this  mean flow by $U_s\equiv U- \langle U \rangle_z$ and the
corresponding streak energy density as
\begin{equation}
\label{eqn:rms_streak}
 E_s=\int_{-1}^1  \frac{1}{2} \langle U_s^{2}\rangle_z\,\df y\
\end{equation}
The streamwise mean velocities of the roll structures are obtained from $V$ and $W$  and  the roll energy density is defined as
\begin{equation}
\label{eqn:rms_roll}
E_r=\int_{-1}^1 \frac{1}{2}\langle V^2+W^2\rangle_z\,\df y.
\end{equation}

The energy of the incoherent motions is determined by the perturbation energy
\begin{equation}
\label{eqn:rms_peturb}
E_p=\int_{-1}^1\,\frac{1}{2}\langle ||\u||^2 \rangle_{x,z}\,\df y.
\end{equation}
The perturbation or streamwise averaged Reynolds stress components are here defined as $\tau_{ij}
\equiv\langle u_i (\xv, t) u_j(\xv,t) \rangle_x \equiv C_{ij}(1,1)$. 

The external excitation $\fv$ is assumed to  be a   temporally white noise process with zero mean satisfying
\be
\label{eq:430}
\langle f_i(\xv_1,t_1) f_j(\xv_2,t_2) \rangle_\infty = \delta(t_1-t_2) Q_{i j } (1,2)~,
\ee
where $\langle \cdot \rangle_N $ indicates an ensemble average over  $N$ forcing realizations.
The ergodic hypothesis is invoked to equate the ensemble mean,  $\langle \cdot \rangle_\infty$, with the
streamwise average, $\langle \cdot \rangle_x$.
 $\Qm(1,2)$ is the matrix covariance between points $\mathbf{x}_1$ and $\mathbf{x}_2$. We assume that
$\Qm(1,2)$ is homogeneous in both $x$ and $z$, i.e. it is  invariant to translations in $x$ and $z$ and therefore has the form: $\Qm(x_1-x_2,y_1,y_2,z_1-z_2)$.

Averaging \eqref{eq:NSEPL} we obtain the equation for the first cumulant:
\begin{equation}
 \label{eq:MNLC}
 \partial_t {\bf{U}} = {\mathbf P_L} \left ( -{\bf{U}} \cdot \nabla {\bf{U}}
 + \frac{1}{Re}\Delta \bf{U} \right ) +  \mathcal{L}\Cm ~.
 \end{equation}

 In this equation  the streamwise average Reynolds stress divergence
$ {\mathbf P_L} ( \left <- \bf{u} \cdot \nabla \bf{u}\right >_x )$,  which
depends linearly on  $\Cm$,  has been
expressed as $\mathcal{L}\Cm$ with $\mathcal{L}$ a linear operator.

 At this point it is important to  notice that the first cumulant  was not set to zero, as is commonly done
 in the study of statistical closures for identifying equilibrium statistical states in isotropic and homogeneous turbulence.   In contrast to the case of isotropic and homogeneous turbulence, retaining the dynamics
of the mean flow, ${\bf U}$,  is of paramount importance in the study of wall-turbulence.

 The second cumulant equation is obtained by  differentiating $C_{i j}(1,2) =   \left < {u}_i (\xv_1)  {u}_j (\xv_2)\right >_x$
with respect to time and using  the equations for the perturbation velocities:
\begin{equation}
\label{eq:pert}
 \partial_t {{\u}} = \A({\bf U})  \u  + \fv  -
 {\mathbf P_L} \left ( \mathbf{u}\cdot \nabla {\u} -  \left < \mathbf{u} \cdot \nabla {\u} \right >_x \right ),
 \end{equation}
Under the ergodic assumption that streamwise averages are equal to ensemble means we obtain:
  \begin{equation}
   \label{eq:CG}
   \partial_t C_{i j}(1,2)= A_{i k}(1) C_{k  j}(1,2)  + A_{ j k}(2) C_{i k}(2,1) +  Q_{i j} (1,2) + G_{i j} ~.
  \end{equation}
In the above   $\A({\bf U})$ is the linearized operator governing evolution of perturbations about the instantaneous mean flow, ${\bf U}$:
  \begin{equation}
  \label{eq:A}
  A({\bf U})_{ij} u_j = {\mathbf P_L} \left ( -{\bf{U} \cdot \nabla} u_i -  {\bf{u} \cdot \nabla } U_i  + \frac{1}{Re}\Delta u_i \right ).
\end{equation}
Notation $A_{i k}( 1)   C_{k j}(1,2)$
indicates that operator $\A$ operates on the velocity variable  of $\Cm$ at position 1, and similarly for $A_{j k} ( 2)  C_{i k}(2,1)$.
The term ${\bf G}$ is proportional to the third cumulant so that  the dynamics of the second
cumulant is not closed.

The first statistical state dynamics we wish to describe is referred to as the stochastic structural stability theory (S3T) system
and it  is obtained by closing the cumulant
expansion at second-order either by assuming that the third cumulant term ${\bf G}$ in~\eqref{eq:CG} is proportional  to  a state independent
covariance homogeneous in $x$ and $z$
 or by setting the third cumulant  to zero.  The former
 is equivalent to parameterizing the term in \eqref{eq:pert}, $
 {\mathbf P_L} \left ( {\bf u} \cdot \nabla \u -  \left < {\bf u} \cdot \nabla { \bf u} \right >_x  \right )$
representing the perturbation-perturbation interactions
by a  stochastic excitation.  This implies that the perturbation dynamics evolve
according to:
\begin{equation}
 \label{eq:FNL}
 \partial_t {\bf{u}} =  \A({\bf U}) \u
  +  \sqrt{\varepsilon}~\fv,
 \end{equation}
 in which the stochastic term $\sqrt{\varepsilon} ~\fv(\xv,t)$,  with  spatial covariance $\varepsilon \Qm(1,2)$  (cf.  \eqref{eq:430}),
 parameterizes the endogenous third order cumulant  in addition to  the exogenous external stochastic excitation, and
 $\varepsilon$ is  a scaling parameter.
 The covariance $\Qm$ can be normalized in energy so that $\varepsilon$ is a parameter indicating the amplitude of the stochastic excitation.
Equation \eqref{eq:MNLC} and  \eqref{eq:FNL}
define what will be referred to as the restricted nonlinear (RNL) dynamics.
Under this parameterization the perturbation nonlinearity responsible for  the turbulent cascade in streamwise Fourier space has been eliminated.
The stochastic structural stability theory (S3T) system is consequently:
\begin{subequations}
\label{eq:RNLc}
 \begin{eqnarray}
 &&\partial_t {\bf{U}} ={\mathbf P_L} \left ( - {\bf U}\cdot \nabla {\bf U}
 +  \frac{1}{Re}\Delta {\bf U} \right ) +  \mathcal{L}\Cm ~~,\label{eqn:m01a} \\
&&\partial_t C_{i j}(1,2)= A_{i k}(1) C_{ k  j}(1,2)  + A_{j k}(2) C_{i k}(1,2) +\varepsilon Q_{i j} (1,2). \label{eqn:p01a}
  \end{eqnarray}
  \end{subequations}
 This is the ideal statistical state dynamics dynamics (SSD)  for studying wall-turbulence using second-order SSD.

Given that the full covariance evolution equation becomes  too large to be directly integrated as the dimension of the dynamics rises with Reynolds number,  a finite number of realizations, $N,$
can be used to  approximate the exact covariance evolution which results {\color{black}  in the  $N$  member ensemble
restricted nonlinear system (RNL$_N$)}:
\begin{subequations}
\label{eq:RNL}
\begin{eqnarray}
\partial_t {{\bf U}} &=&{\mathbf P_L} \left ( - {\bf U}\cdot \nabla {\bf U}
 +  \frac{1}{Re}\Delta {\bf U} - \left <  \left  <\bf{u} \cdot \nabla \bf{u}\right >_x  \right >_N\right)  ~~,\label{eqn:RNL-mean} \\
 \partial_t {\u_n} &= &\A({\bf U}) {\u_n} +
  \sqrt{\varepsilon}~{\fv_n}, ~~(n=1,\dots,N). \label{eqn:RNL-pert}
\end{eqnarray}
  \end{subequations}
%
The average  $\langle \cdot \rangle_N$ in \eqref{eqn:RNL-mean} is obtained using an $N$-member ensemble of  realizations of  \eqref{eqn:RNL-pert} each of which results from a statistically independent stochastic excitation
$\fv_n$ but in which all share the same ${\bf U}$.  When an infinite ensemble is used the RNL$_\infty$ system is obtained which is   equivalent to  the S3T system \eqref{eq:RNLc}.
Remarkably, a single ensemble member often suffices to obtain a useful approximation to the covariance evolution, albeit with substantial statistical fluctuations.  In the case $N=1$, equation \eqref{eq:RNL} can be viewed as both an approximation of statistical state dynamics (SSD) and a realization of RNL dynamics.  When $N>1$ it is only an approximation to the SSD.

\section{Using S3T to obtain analytical solutions for turbulent  states}
\label{sec:S3T}
\begin{figure*}
	\centering
       \includegraphics[width = 1.\textwidth]{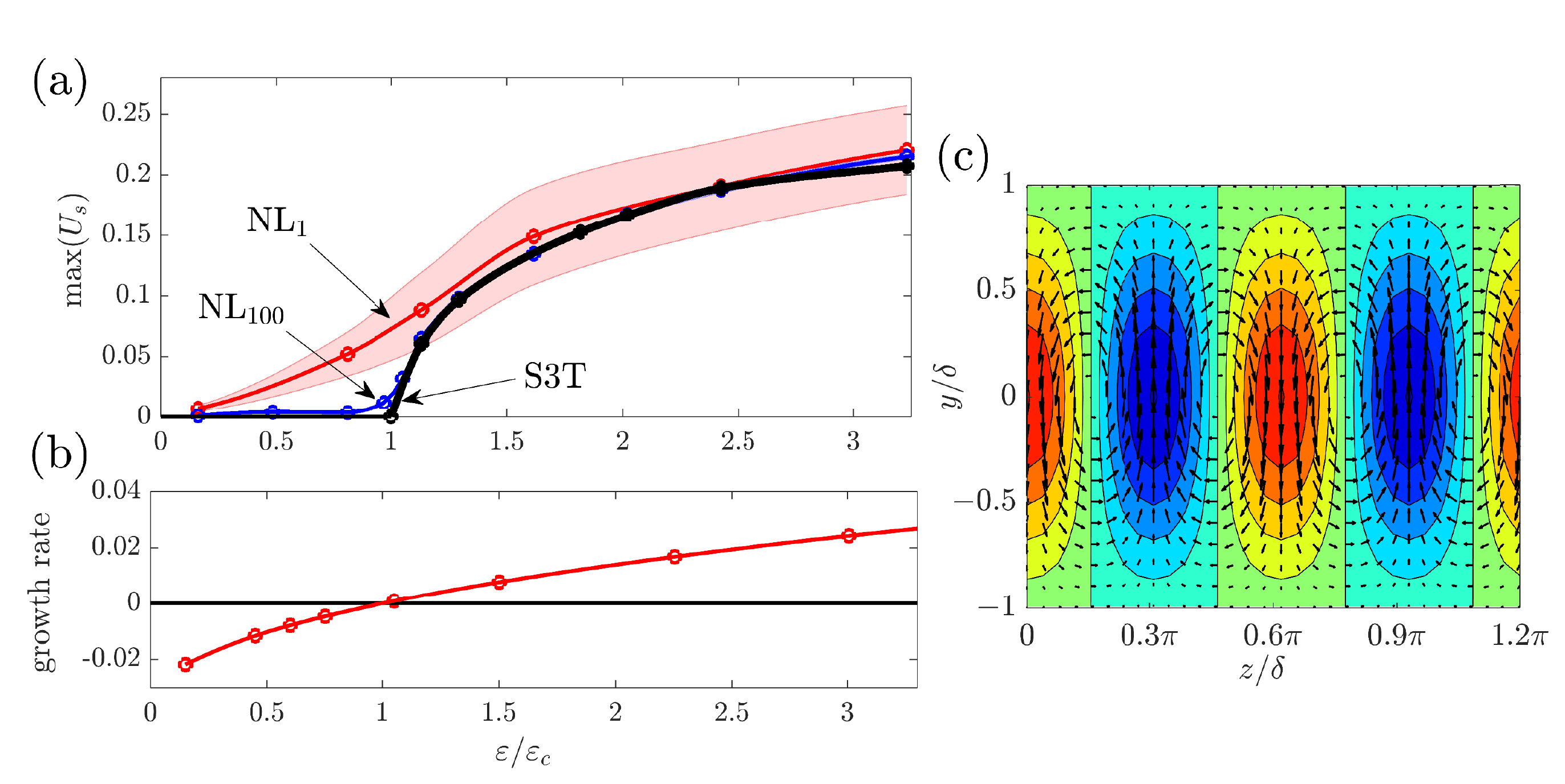}
        \caption{Analysis of  roll/streak formation  from an statistical state dynamics (SSD) bifurcation in a minimal channel Couette flow forced by background turbulence.
         Panel (a): streak amplitude, $U_s$, as a function of the stochastic excitation  amplitude, $\varepsilon$, revealing the
bifurcation as predicted by S3T (black)
and the reflection of this prediction in  an NL$_1$ simulation (red)  and in  an NL$_{100}$
simulation (blue).  The NL$_1$   simulations  exhibit fluctuations from the analytical predicted roll/streak structure
with one standard deviation of the fluctuations indicated by shading.  The critical value $\varepsilon_c$ is obtained from S3T stability analysis of the spanwise homogeneous state.
 The underlying S3T  eigenmode
is shown in  panel (b) and  its growth rate in (c).
In panel (b) streak velocity, $U_s$,  is indicated by contours and the velocity components $(V, W)$ by vectors.
At  $\varepsilon=\varepsilon_c$  the S3T spanwise uniform equilibrium bifurcates to a finite amplitude equilibrium
with perturbation structure close to that of the most unstable eigenfunction  shown in (b).
The channel is minimal with $L_x=1.75 \upi$ and $L_z=1.2 \upi$  \cite{Jimenez-Moin-1991},
the Reynolds number is $Re=400$, and the stochastic forcing excites
only  Fourier components with  streamwise wave number $k_x= 2 \upi/L_z= 1.143$.
 Numerical calculations  employ $N_y=21$ grid points in the cross-stream direction and
$32$ harmonics in the spanwise and streamwise directions (Adapted from \cite{Farrell-Ioannou-2016-bifur}).}
\label{fig:bifur_en}
\end{figure*}
Streamwise roll vortices and associated streamwise streaks are  prominent features in transitional boundary layers ~\cite{Klebanoff-etal-1962}.   The ubiquity
of the roll/streak structure in these flows presents a problem because the laminar solution of these flows is linearly stable. 
However, because of the high non-normality of the Navier-Stokes (NS)
dynamics linearized about a strongly sheared flow streamwise constant structures such the roll/streak have the greatest transient growth providing an explanation for its arising from perturbations to the flow \cite{Butler-Farrell-1992, Reddy-Henningson-1993}.  However,   stochastic structural stability theory (S3T) reveals that
 the roll/streak structure is destabilized by systematic organization by the streak  of the perturbation Reynolds stress associated with
 low levels of {\color{black}  background  turbulence} \cite{Farrell-Ioannou-2012}.
Destabilization
of the roll/streak  can be traced to a universal positive feedback mechanism operating in turbulent flows:  the coherent streak distorts the incoherent turbulence so as to induce ensemble mean  perturbation Reynolds stresses that force streamwise mean roll circulations configured to reinforce the streak  (cf. \cite{Farrell-Ioannou-2012}).  The modal streak perturbations of the fastest growing eigenfunctions  induce the strongest such feedback.
This instability does not have analytical expression in eigenanalysis of  the NS dynamics  but it can be solved for by performing an eigenanalysis on the S3T system.

Consider a laminar plane Couette flow  subjected to stochastic excitation that is  statistically streamwise and spanwise homogeneous and has zero spatial and temporal mean.    S3T predicts that a bifurcation occurs at a critical amplitude of excitation, $\varepsilon_c$, in which an unstable  mode with roll/streak structure emerges ($\varepsilon_c$ corresponds to an energy input rate that would sustain
background  turbulence energy of $0.14\%$ of the laminar flow).
As the  excitation parameter, $\varepsilon$ in \eqref{eqn:p01a},  is increased  finite amplitude roll/streak structures equilibrate from this instability \cite{Farrell-Ioannou-2012}.  While these  equilibria underlie the dynamics of roll/streak formation in the pre-transitional flow,  they are imperfectly reflected in individual realizations  (cf. Refs.~\cite{Farrell-Ioannou-2003-structural,Farrell-Ioannou-2016-bifur}).    One can compare this behavior to that  of the corresponding  Navier-Stokes (NS) solutions by {\color{black} defining the  $N$ ensemble nonlinear  system (NL$_N$) } in
analogy with RNL$_N$ as follows:
\begin{subequations}
\label{eq:eDNS}
\begin{gather}
\partial_t {{\bf U}} ={\mathbf P_L} \left ( - {\bf U}\cdot \nabla {\bf U}
 +  \frac{1}{Re}\Delta {\bf U} - \left <  \left  <\bf{u} \cdot \nabla \bf{u}\right >_x  \right >_N\right)  ~~,\label{eq:eDNSm} \\
%
 \partial_t {{\u_n}} = \A({\bf U})  \u_n   +\sqrt{\varepsilon} ~ \fv_n  -
 {\mathbf P_L} \left ( \u_n \cdot \nabla {\u_n} -  \left < \u_n \cdot \nabla {\u_n} \right >_x  \right ),  ~~(n=1,\dots,N) \label{eq:eDNSp}~.
\end{gather}\label{eq:eNSE00}\end{subequations}
 Note that as  $N \to \infty$ this system provides  the second-order
 statistical state dynamics of the Navier-Stokes (NS) without approximation.  Fig. \ref{fig:bifur_en} compares the analytical bifurcation structure predicted by S3T,  the quasi-equilibria obtained using a single realization of  the Navier-Stokes  (NL$_1$)  and the near perfect reflection of the
 S3T bifurcation  in a $100$ member  Navier-Stokes ensemble (NL$_{100}$)
 (cf.  Refs. \cite{Farrell-Ioannou-2012, Farrell-Ioannou-2016-bifur}).

With continued increase in $\varepsilon$  a second bifurcation occurs in which the flow transitions to a chaotic time-dependent state.
 For the parameters used in our example this second bifurcation occurs at $\varepsilon_t/\varepsilon_c=5.5$.
Once this time-dependent state is established the stochastic forcing can be removed and this state continues to be maintained as a self-sustaining turbulence.  Remarkably,  this  self-sustaining turbulence naturally simplifies further by evolving to a minimal turbulent system in which the dynamics is supported by the interaction of the roll-streak structures with a perturbation field comprising a small number of streamwise harmonics (as few as 1). This minimal self-sustaining turbulent system, which proceeds naturally from the S3T dynamics, reveals an underlying self-sustaining process (SSP) which can be understood with clarity. The basic ingredient of this SSP is the robust tendency for streaks to organize the perturbation field so as to produce streamwise Reynolds stresses supporting the streak,  as in the S3T instability mechanism  shown  Fig. \ref{fig:bifur_en}c. Although the streak is strongly fluctuating in the self-sustaining state, the tendency of the streak to organize the perturbation field is retained. It is remarkable that the perturbations, in this highly time-dependent state, produce torques that maintain the streamwise roll not only on average but at nearly every instant. As a result, in this self-sustaining state, the streamwise roll is systematically maintained by the robust organization of perturbation Reynolds stress by the time-dependent streak while the streak is maintained by the streamwise roll through the lift-up mechanism \cite{Farrell-Ioannou-2012, Constantinou-etal-2014b}. Through the resulting time-dependence of the roll-streak structure the constraint on instability imposed by the absence of inflectional instability in the mean flow is bypassed as the perturbation field is maintained by parametric growth \cite{Farrell-Ioannou-2012,Farrell-Ioannou-2016-sync}.

\section{Self-sustaining turbulence in a restricted nonlinear model}
\label{sec:RNL_selfsustain}
The {\color{black} previous sections demonstrated that the  stochastic structural stability theory  (S3T) system    \eqref {eq:RNLc}}  provides an attractive theoretical framework for studying turbulence through analysis of its underlying statistical mean state dynamics. However, it has the perturbation covariance as a variable and its dimension, which is O($N^2$) for a system of dimension O($N$), means that
it is directly integrable only for low order systems. {\color{black} In this section} we demonstrate that this computational limitation can be overcome by instead simulating the {\color{black} $N$  ensemble member RNL$_N$  \eqref{eq:RNL},  using a finite
number of realizations of the perturbation field \eqref{eqn:RNL-pert}}.  {\color{black}In particular, we perform computations for a plane Couette flow at $Re=1000$, which }  show
that  a single realization ($N=1$) suffices to approximate the ensemble covariance allowing computationally efficient studies of the dynamical restriction underlying the S3T dynamics. We then demonstrate that the
{\color{black} single  ensemble member RNL$_1$ (which we interchangeably refer to
as the RNL system) }reproduces self-sustaining turbulent dynamics that reproduce the key features of turbulent plane Couette flow at low Reynolds numbers.  We show that in correspondence with the S3T results, RNL turbulence is supported by a perturbation field comprising only a few streamwise varying modes (harmonics or $k_x\neq0$ Fourier components in a Fourier representation) and that its streamwise wave number support can be reduced to a single streamwise varying mode interacting with the streamwise constant mean flow.
\begin{figure}
  \centering\hspace{-0.2in}
     \subfigure[{\bf \ S3T} \label{fig:S3T_contour}]{\hspace{-0.215in}
  \includegraphics[width = 0.352\textwidth,clip]{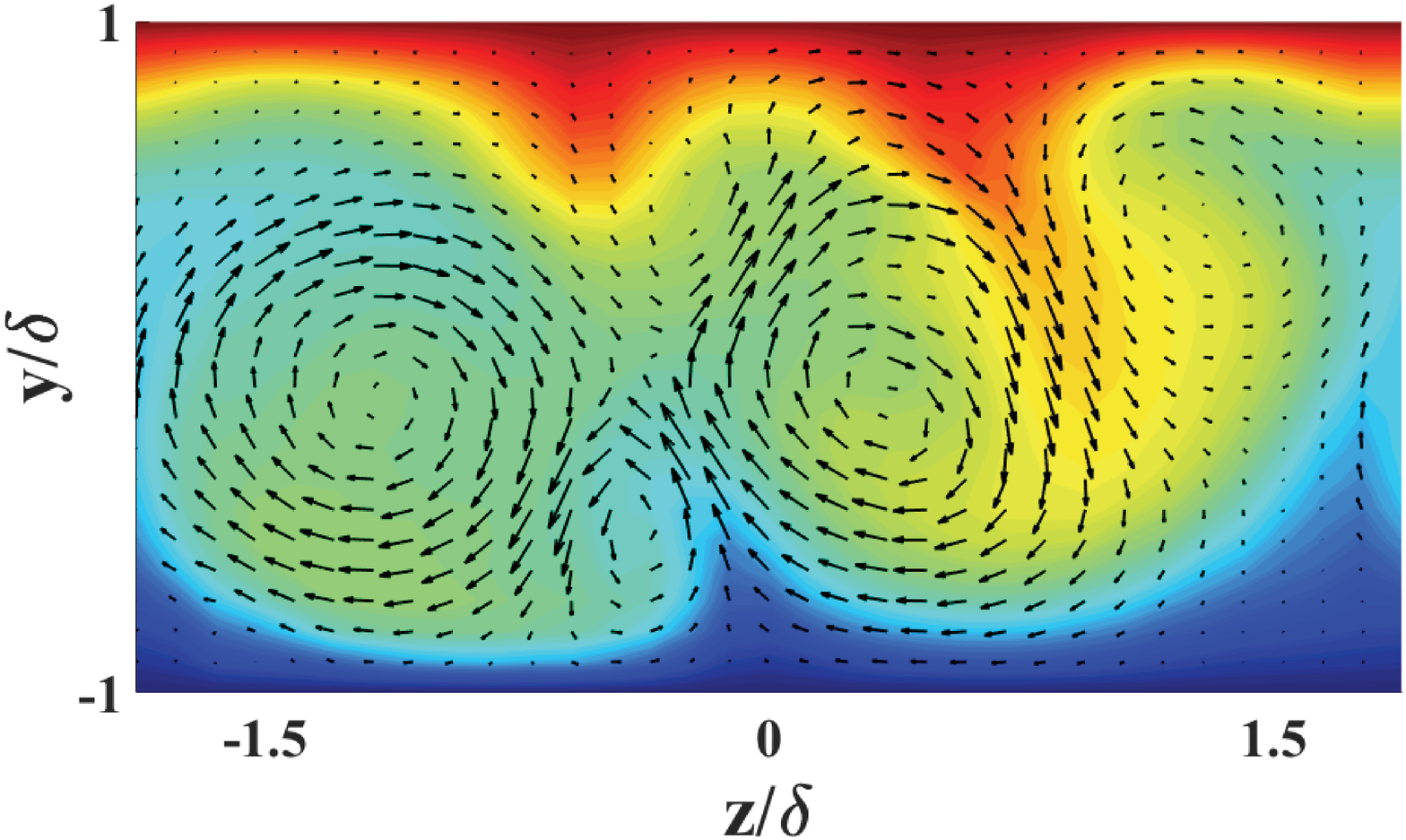}
   }
  \subfigure[{\bf \ RNL} \label{fig:RNL_contour}]{\hspace{-0.215in}
  \includegraphics[width = 0.352\textwidth,clip]{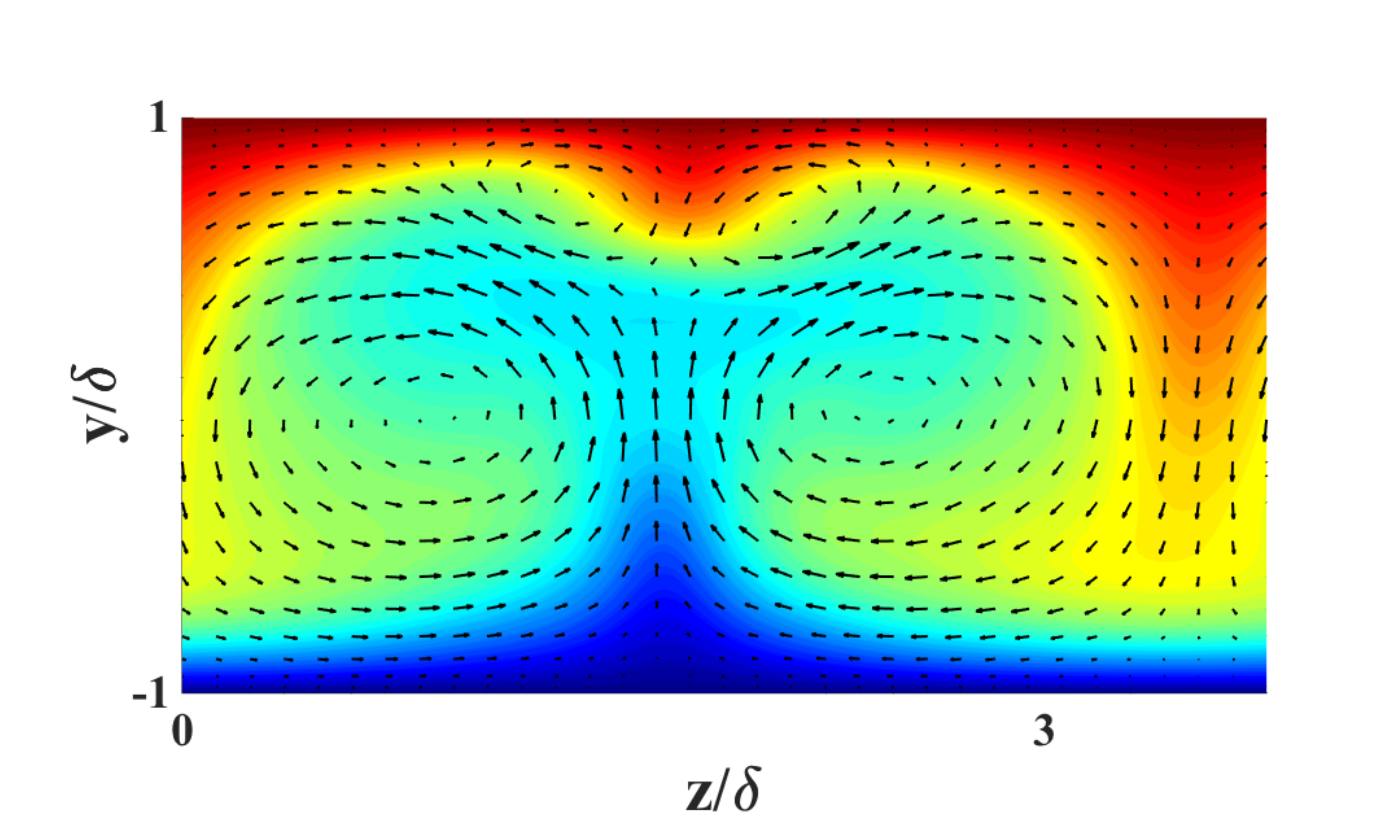}
  }
\subfigure[{\bf \ DNS} \label{fig:DNS_contour}]{\hspace{-0.15in}
   \includegraphics[width = 0.352\textwidth,clip]{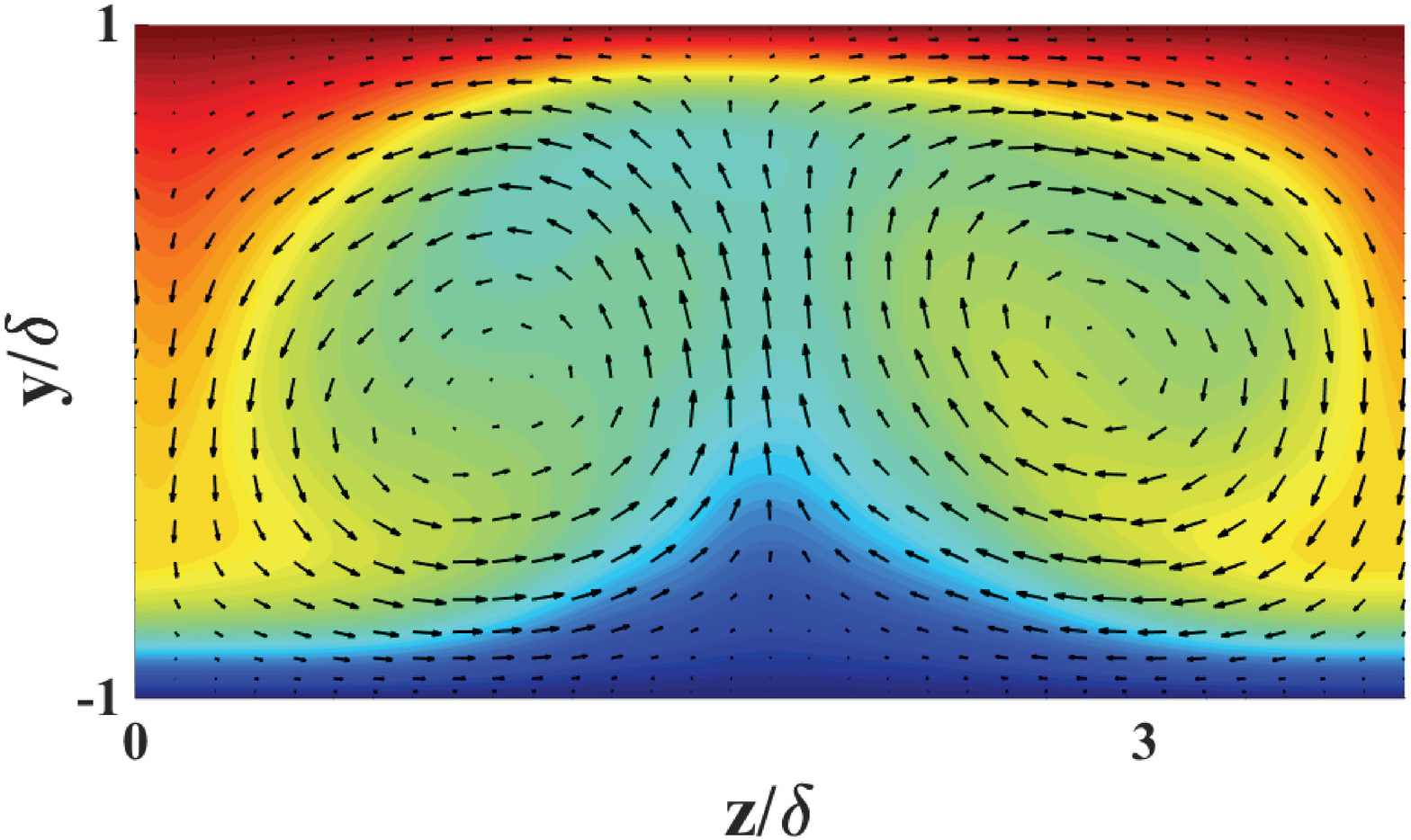}
   }
   \caption{A $y$-$z$ plane cross-section of the flow (at $x$ = 0) at a single snapshot in time
   for (a) {\color{black}a S3T simulation, (b) a RNL simulation and (c) DNS data}. All panels show contours of the
   streamwise component of the mean flow $U$ with the velocity vectors $V$, $W$ superimposed. The RNL and S3T dynamics are self-sustaining for the time shown.  }\vspace{-20pt}
\label{fig:contours}
\end{figure}

{\color{black} We initiate turbulence in all of the RNL plane Couette flow simulations in this section} by applying a stochastic
excitation $\mathbf{f}$ in \eqref{eqn:RNL-pert} over the  interval $t \in [0,500]$.
We apply a similar procedure to initiate turbulence in the DNS, through $\mathbf{f}$ in \eqref{eq:NSE0}, and S3T simulations, through its spatial covariance $\mathbf{Q} (1,2)$ in \eqref{eqn:p01a}.  All results reported are for $t>1000$, unless otherwise stated. The DNS results are obtained from the Channelflow NS solver \cite{channelflow,Gibson-etal-2008}, which is a pseudospectral code. The RNL simulations use a modified version of the same code. Complete details are provided in \cite{Thomas2014}.

A comparison of the velocity field obtained from S3T and RNL$_1$ simulations that have reached self-sustaining states (i.e. for is $t>1000$) is shown in figures \ref{fig:S3T_contour} and \ref{fig:RNL_contour}. These panels depict contour plots of an instantaneous snapshot of the streamwise component of the mean velocity with the vectors indicating velocity components ($V$, $W$) superimposed for the respective S3T and RNL flows at {\color{black}$Re=600$} in a minimal channel, see the caption in Figure  \ref{fig:bifur_en} for the details.  The same contour plot for a DNS is provided in figure  \ref{fig:DNS_contour} for comparison. These plots demonstrate the qualitative similarity in the structural features obtained from an S3T simulation, where the mean flow is driven by the full covariance, and the RNL simulation in which the covariance is approximated
with a single realization of the perturbation field.  Both flows also show  good qualitative agreement with the DNS data.

\begin{figure}
\subfigure[\label{fig:TurbulentProfile}]{
\includegraphics[width = 0.45\textwidth,clip=]{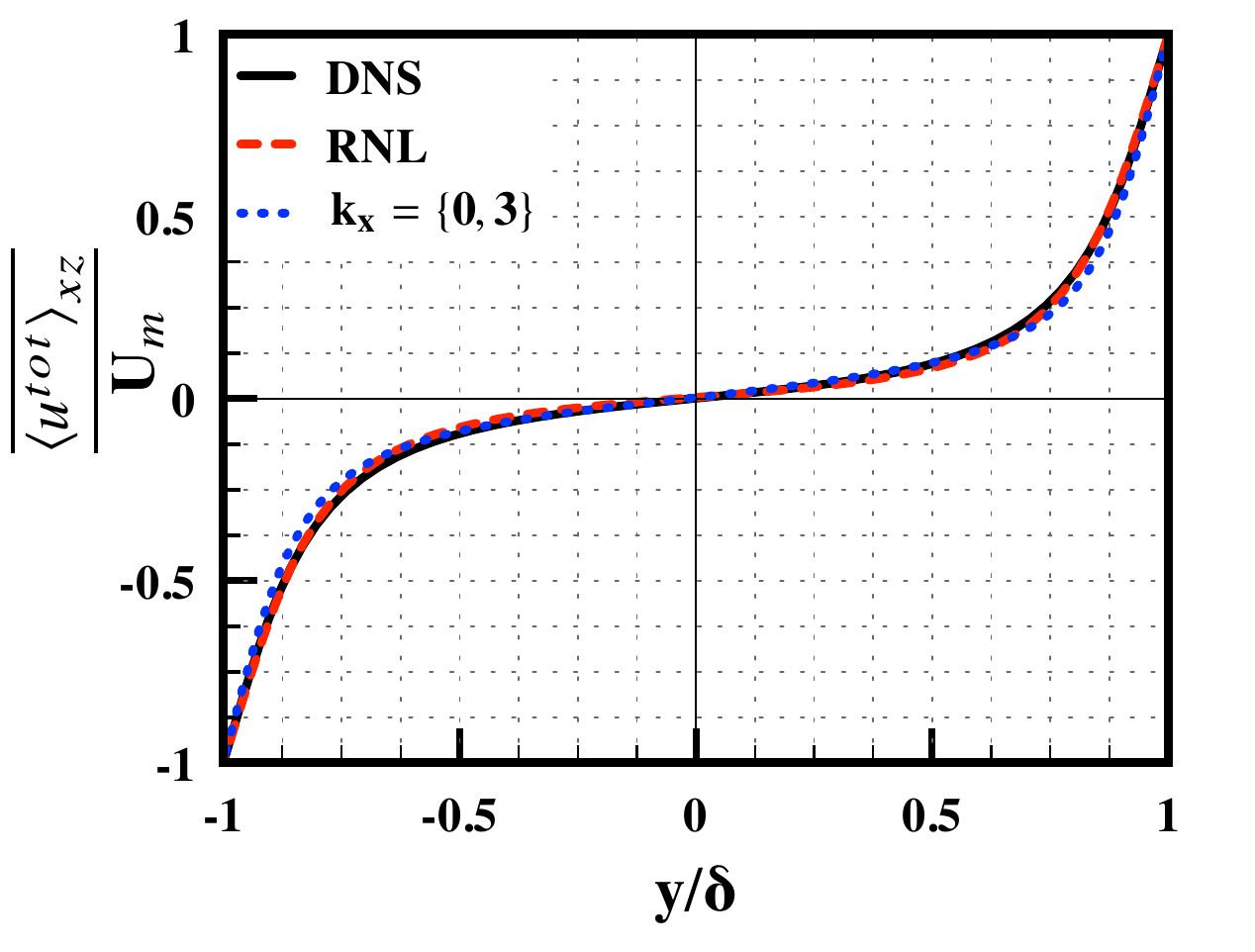}}
\subfigure[\label{fig:Reynolds}]{
\includegraphics[width = 0.45\textwidth,clip=]{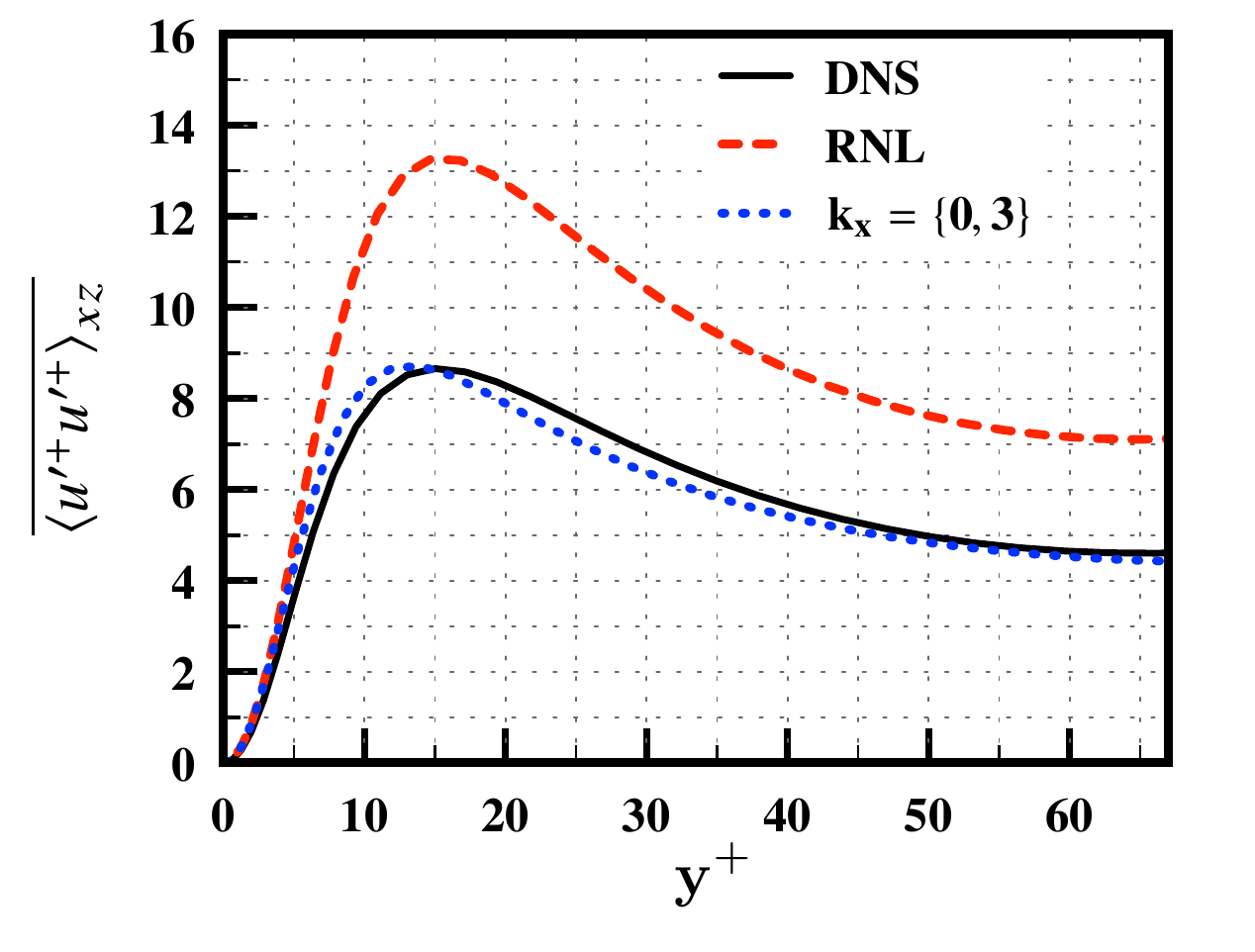} }
\caption{(a) Turbulent mean velocity profiles (based on streamwise, spanwise and time averages) in geometric units  obtained from the DNS (red solid line) and RNL simulations  with no band-limiting (black dashed line) and one where the streamwise wave number support is limited to $k_x=[0,3]$ (blue dotted line).  All at $R=1000$. (b) Reynolds stresses $\overline{\langle u'^+u'^+}\rangle _{xz}$, where {\color{black}$u'$ =  $u^{tot} - \overline{u^{tot} }$} and $u^+$ = $u/u_{\tau}$. The figure in panel (a) is adapted from \cite{Thomas2014}.}
  \label{fig:turbulentprofiles}
 \end{figure}
Having established the ability of the RNL system to provide a good qualitative approximation of the S3T turbulent field, we now proceed to discuss the features of RNL turbulence.  For this discussion we move away from the minimal box at {\color{black}$Re=600$} that was used to facilitate comparison with the S3T equations and instead study plane Couette flow at $Re=1000$ in a box with respective streamwise and spanwise extents of $L_x=4\pi\delta$ and $L_z=4\pi\delta$.  The turbulent mean velocity profile obtained from a DNS and a RNL simulation under these conditions is shown in Figure \ref{fig:TurbulentProfile}, which illustrates good agreement between the two turbulent mean velocity profiles. Figure \ref{fig:Reynolds} shows the corresponding time-averaged  Reynolds stress component, $\overline{\langle u'^+u'^+}\rangle _{xz}$,  where the streamwise fluctuations, $u'$, are defined as {\color{black}$u'$ =  $u^{tot} - \overline{u^{tot} }$,  $u^+$ = $u/u_{\tau}$ and $y^+$ = $(y+1) \hspace*{0.3 mm} u_{\tau}/\nu$ with friction velocity  $u_{\tau}$ = $\sqrt{\tau_w/\rho}$,  (where $\tau_{w}$ is the shear stress at the wall), $Re_{\tau}$ = $u_{\tau}\delta/\nu$ and $\nu$ = $1/Re$.  The friction Reynolds numbers for the DNS data and the RNL simulation are respectively,  ${Re_{\tau}}$ = $66.2$ and  ${Re_{\tau}}$ = $64.9$. {\color{black} Although they are not shown here, previous studies have also shown close agreement between the Reynolds shear stress $\overline{u'v'}$ obtained from the RNL simulation and DNS \cite{Thomas2014}, which is consistent with the fact that the turbulent flow supported by DNS and the RNL simulation exhibit nearly identical shear at the boundary, as seen in Figure \ref{fig:TurbulentProfile}.
The close correspondence in the mean profiles in Figure \ref{fig:TurbulentProfile}, the $\overline{u'v'}$ Reynolds stresses reported in \cite{Thomas2014},  as well as in the close correspondence values of $Re_\tau$ in the RNL simulations and DNS  indicate that the overall energy dissipation rates per unit mass $\mathcal{E}=\tau_{w}{U}/{\delta}$, where ${U}/{\delta}$ is a constant based on the velocity of the walls $U$ and the half-height of the channel $\delta$, also show close correspondence}.

  Figure \ref{fig:Reynolds} shows} that the peak magnitude of the streamwise component of the time-averaged Reynolds stresses, $\overline{\left<u'^+u'^+\right>_{xz}}$, is too high in the RNL simulation.  {\color{black} Other second order statistics the premultiplied streamwise and spanwise spectra for this particular flow are presented in \cite{Thomas2014}.  The discrepancies in both $\overline{\left<u'^+u'^+\right>_{xz}}$ and the streamwise premultiplied spectra reported are} a direct result of the dynamical restriction, which  results in a reduced number of streamwise wave numbers that support RNL turbulence, which we discuss next.  In particular, we demonstrate that when $\mathbf{f}$ in equation \eqref{eqn:RNL-pert} is set to 0, the RNL model reduces to a minimal representation in which only a finite number of streamwise varying perturbations are maintained while energy in the other streamwise varying perturbations decays exponentially.  {\color{black} This resulting limited streamwise wave number support cannot and is not expected to accurately reproduce the entire streamwise spectra but instead captures the spectral components associated with the turbulent structures that are responsible for the self-sustaining process,  i.e. those corresponding to  the spanwise streak and roll.}


In order to frame our discussion of the streamwise wave number support of RNL turbulence we define a streamwise energy density associated with each perturbation wave number $k_n, \, (n\neq 0)$ based on the perturbation energy of the associated streamwise wavelength $\lambda_n$ as
    \begin{equation}
    E_{\lambda_n}(t) =  \int_{-1}^{1}  \frac{1}{4}\langle  ||\mathbf{u}_{\lambda_n}(y,z,t)||^2  \rangle_z  \ dy.
    \end{equation}
Here $\mathbf{u}_{\lambda_n}$ is the perturbation, $\mathbf{u}=(u,v,w)$, associated with Fourier components with streamwise wavelength $\lambda_n$. We refer to the set of streamwise wave numbers for which    $E_{\lambda_n}(t)$  does not tend to zero when $\mathbf{f}=0$ in equation \eqref{eqn:RNL-pert}, as the  natural support for the RNL system.

\begin{figure}[t]
\centering
\subfigure[\label{fig:RNL_support}]{
\includegraphics[width = 0.45\textwidth,clip=]{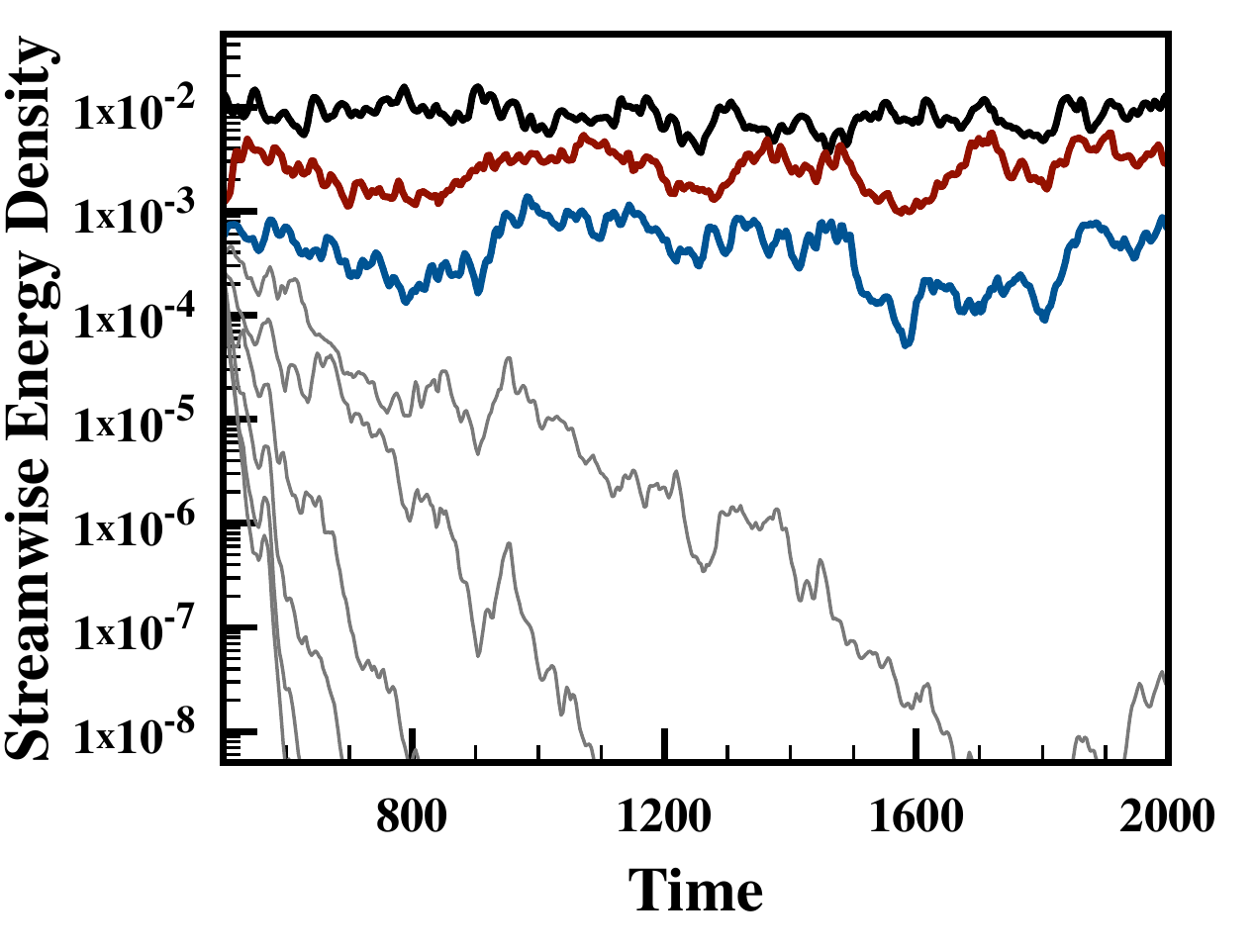}}
\subfigure[\label{fig:DNS_support}]{
\includegraphics[width = 0.45\textwidth,clip=]{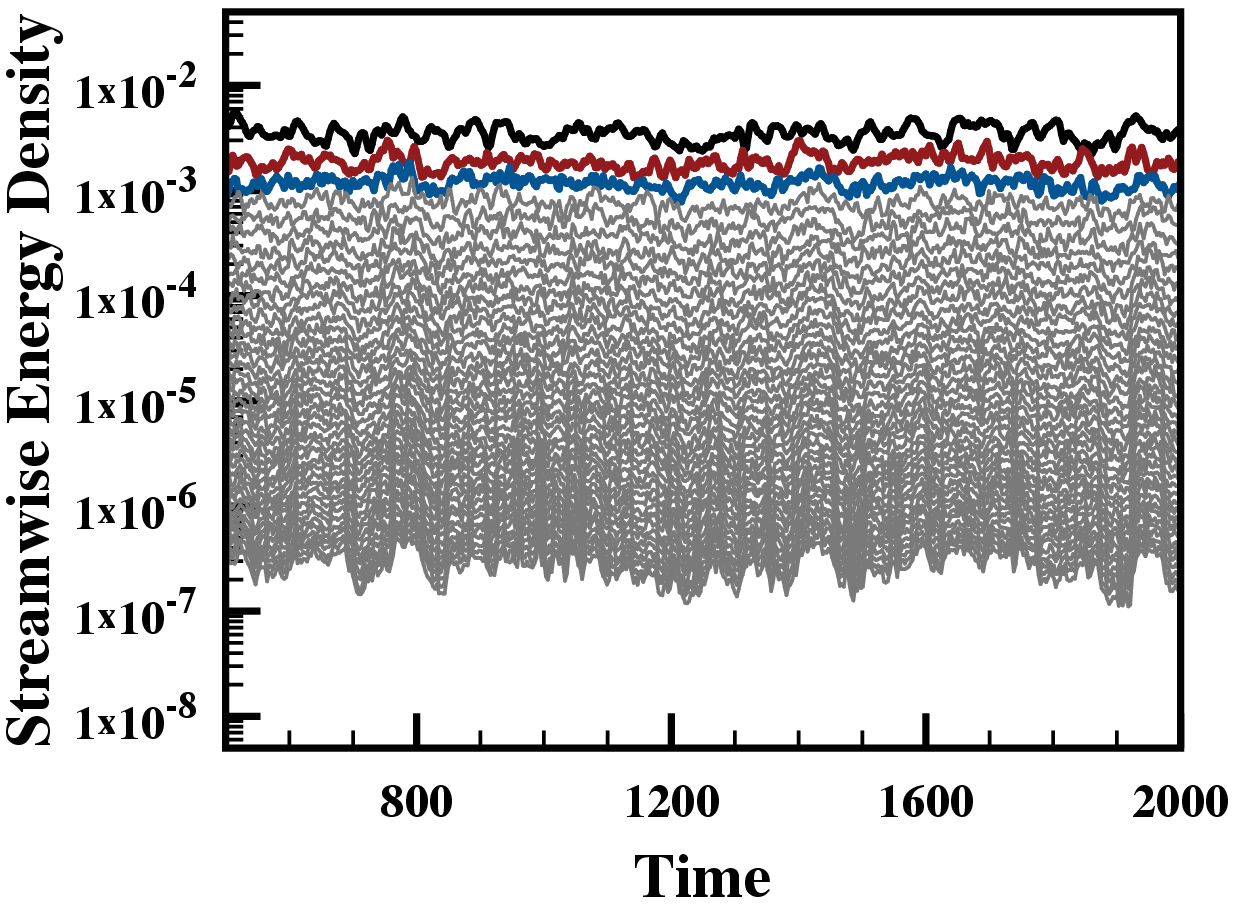}}
\caption{Streamwise energy densities for (a) a RNL simulation and (b) a DNS starting at $t=500$, when the stochastic excitation was terminated.
 The energy densities of the streamwise varying perturbations that are supported in the RNL simulation are shown in the following manner $\lambda_1$ = 4$\pi \delta$ (black),
$\lambda_2$ = 2$\pi \delta$ (red), $\lambda_3$ = 4$\pi \delta /3$ (blue).  The modes that decay when the RNL is in a self-sustaining state are shown in grey in both panels.}
  \label{fig:natural_support}
 \end{figure}

Figures \ref{fig:RNL_support} and \ref{fig:DNS_support} shows the time evolution of the streamwise energy densities $E_{\lambda_{n}}$ for a DNS and a RNL simulation, respectively.  The simulations were both initiated with a stochastic excitation containing a full range of streamwise and spanwise Fourier components that was applied until $t=500$. Figure  \ref{fig:RNL_support} illustrates that the streamwise energy density of most of the modes in the RNL simulation decay once the stochastic excitation is removed.  The decay of these modes is a result of the dynamical restriction not an externally imposed modal truncation.  As a result, the self-sustaining turbulent behavior illustrated in Figure \ref{fig:turbulentprofiles} is supported by only 3 streamwise varying modes.  In contrast,  all of the perturbation components remain supported in the DNS.
This behavior highlights an appealing reduction in model order in a RNL$_1$ simulation, which is consistent with the order reduction obtained
when $N\to \infty$~\cite{Farrell-Ioannou-2012}.

We now demonstrate that RNL turbulence can be supported even when the perturbation dynamics \eqref{eqn:RNL-pert} are further restricted to a single streamwise Fourier component. This restriction to a particular wave number or set of wave numbers is accomplished by slowly damping the other streamwise
varying modes as described in \cite{Thomas2015}. We refer to a RNL$_1$ system that is truncated to a particular set of streamwise Fourier components as a band-limited RNL model and those with no such restriction as baseline RNL systems.

 \begin{figure}[t]
\centering\hspace{-0.2in}
\subfigure[]{\label{fig:kx3Spectra}
\includegraphics[width = 0.5\textwidth,clip=]{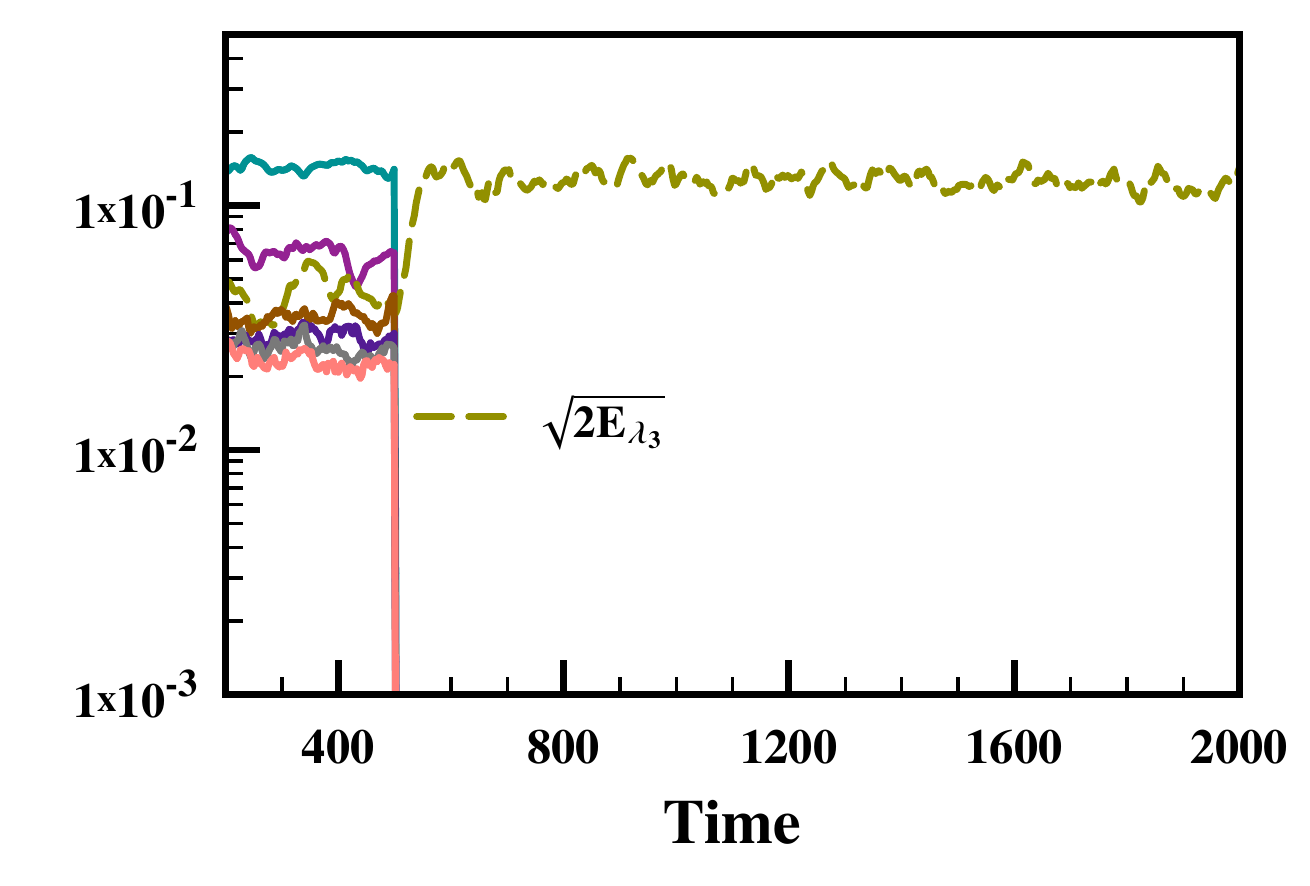}}
\subfigure[]{\label{fig:kx3RMS}
\includegraphics[width = 0.5\textwidth,clip=]{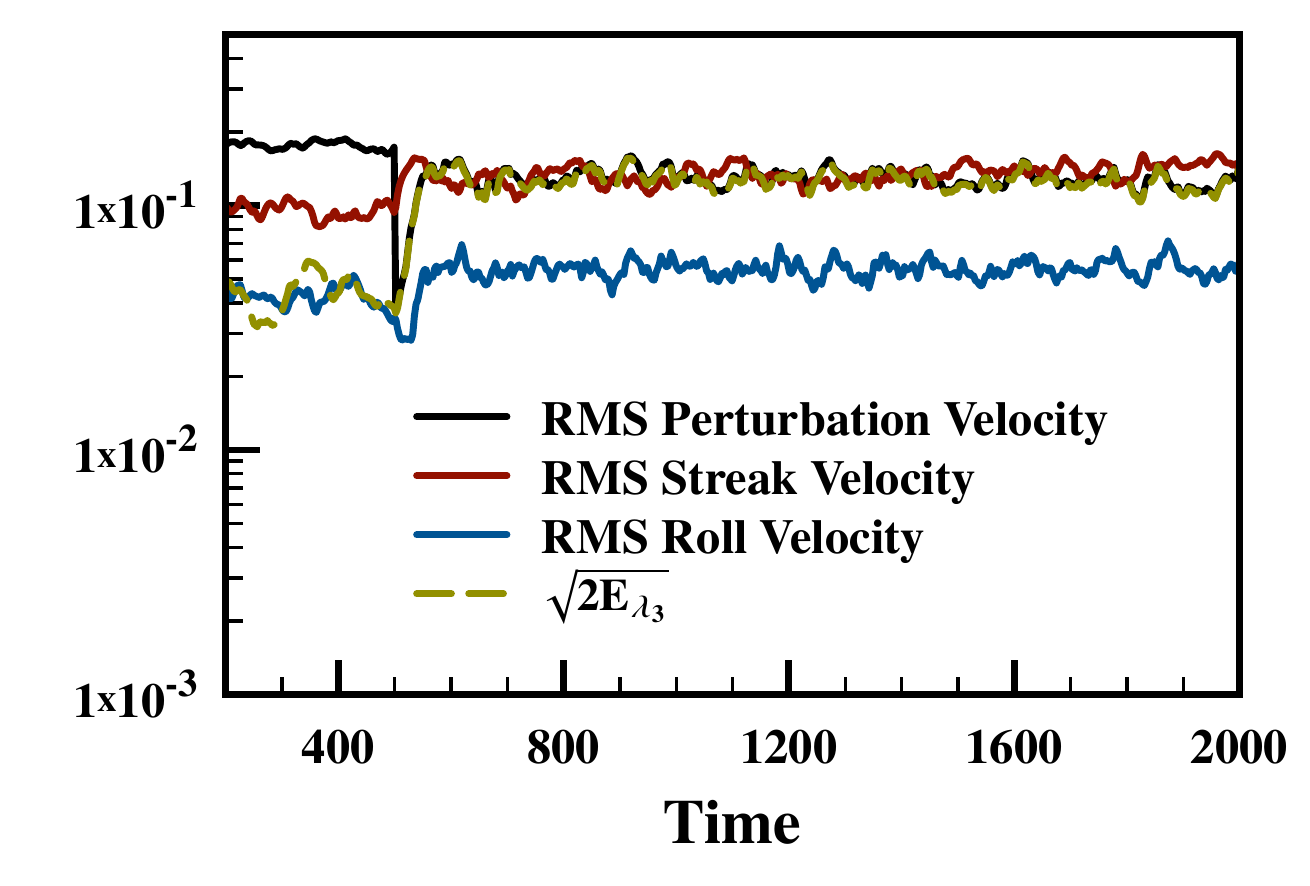}}
\caption{Sustaining turbulence with a single streamwise varying mode. (a) The RMS velocity of the streamwise energy density, $\sqrt{2E_{\lambda_n}}$,  contained in each of the streamwise varying modes versus time before and after broad spectrum forcing is removed at $t=500$. The remaining mode $k_x=3$ ($\lambda_3=4\pi \delta /3$) is shown in gold. After the forcing is removed the remaining mode increases its energy to compensate for the loss of the other modes. (b) $\sqrt{2E_{\lambda_n}}$ for the undamped wavelength along with the RMS perturbation velocity, $U_{pert}$, the RMS streak velocity, $U_{streak}$, and the RMS roll velocity, $U_{roll}$, for the same data as in (a). }
  \label{fig:single_mode}
 \end{figure}

Thomas et al. \cite{Thomas2015} showed that band-limited RNL systems produce
mean profiles and other structural features that are consistent with the baseline RNL system.
Here we discuss only a subset of those results focusing on the particular case in which we keep
only the $k_x=3$ mode corresponding to $\lambda=4\pi/3\delta$.
Figure \ref{fig:kx3Spectra} shows the time evolution of the RMS velocity associated with
the streamwise energy density, $\sqrt{2E_{\lambda_3}}$.  The figure begins just prior
to the removal at $t=500$ of the full spectrum stochastic
forcing used to initialize the turbulence. At $t=500$
all but the perturbations associated with streamwise wave number
$k_x=3$ are removed.  It is interesting to note that
 once these streamwise wave numbers
 are removed the energy density of the remaining mode increases to maintain the turbulent state.  This behavior can be further examined in the evolution of the RMS velocities of the streak, roll and perturbation energies over the same time period, which are respectively defined as $U_{streak}=\sqrt{2 E_s}$, $U_{roll}=\sqrt{2 E_r}$ and $U_{pert}=\sqrt{2 E_p}$, where $E_s$, $E_r$ and $E_p$ are respectively defined in equations \eqref{eqn:rms_streak} - \eqref{eqn:rms_peturb} shown in Figure \ref{fig:single_mode}.
 Here it is clear that after a small transient phase the roll and streak structures supported through the $k_x=3$ perturbation field increase to the levels maintained by the larger number of perturbation components present prior to the band-limiting.

Figure \ref{fig:Reynolds} also shows that the streamwise component of the normal Reynolds stress obtained in this band-limited system shows
better agreement with the DNS than does  the baseline RNL system.  This behavior can be explained by looking at Figure \ref{fig:kx3RMS}, which shows that once the forcing is removed the total perturbation energy (as seen through $U_{pert}$) falls only slightly. This small drop is likely due to the removal of the forcing.  This is consistent with observations that baseline and band-limited RNL simulations have approximately the same perturbation energy.  The lower turbulent kinetic energy in  Figure \ref{fig:Reynolds} for the band-limited system can be attributed to the increase in dissipation that results from forcing the flow to operate with only
the shorter wavelength (higher wave number) structures.

\section{RNL turbulence at moderate Reynolds numbers}
\label{sec:RNL_halfchannel}
The previous section demonstrates that the {\color{black}low order statistics obtained from RNL$_1$ simulations of low Reynolds number plane Couette flow show good agreement with DNS}.  We now discuss how the insight gained at low Reynolds numbers can be applied to simulations of half channel flows at moderate Reynolds numbers.  {\color{black} The half-channel flow equations are given by equation \eqref{eq:NSE0} with a constant pressure gradient $\partial_x p_{\infty} $, a characteristic velocity $U_m$ equal to velocity at the top of the half-channel for the laminar flow, and the characteristic length $\delta$ equal to the full half-channel height. No-slip and stress-free boundary conditions are imposed at the respective bottom and top walls. As in the previous configurations periodic boundary conditions are imposed in the streamwise and spanwise directions. Further details regarding the half-channel simulations}  are provided in \cite{Bretheim-etal-2015}.  All results reported in this section are for $\varepsilon=0$.

\begin{figure*}[!t]
\centering
\subfigure[\label{fig:mvps}]{
\includegraphics[width = 0.5\textwidth,clip=]{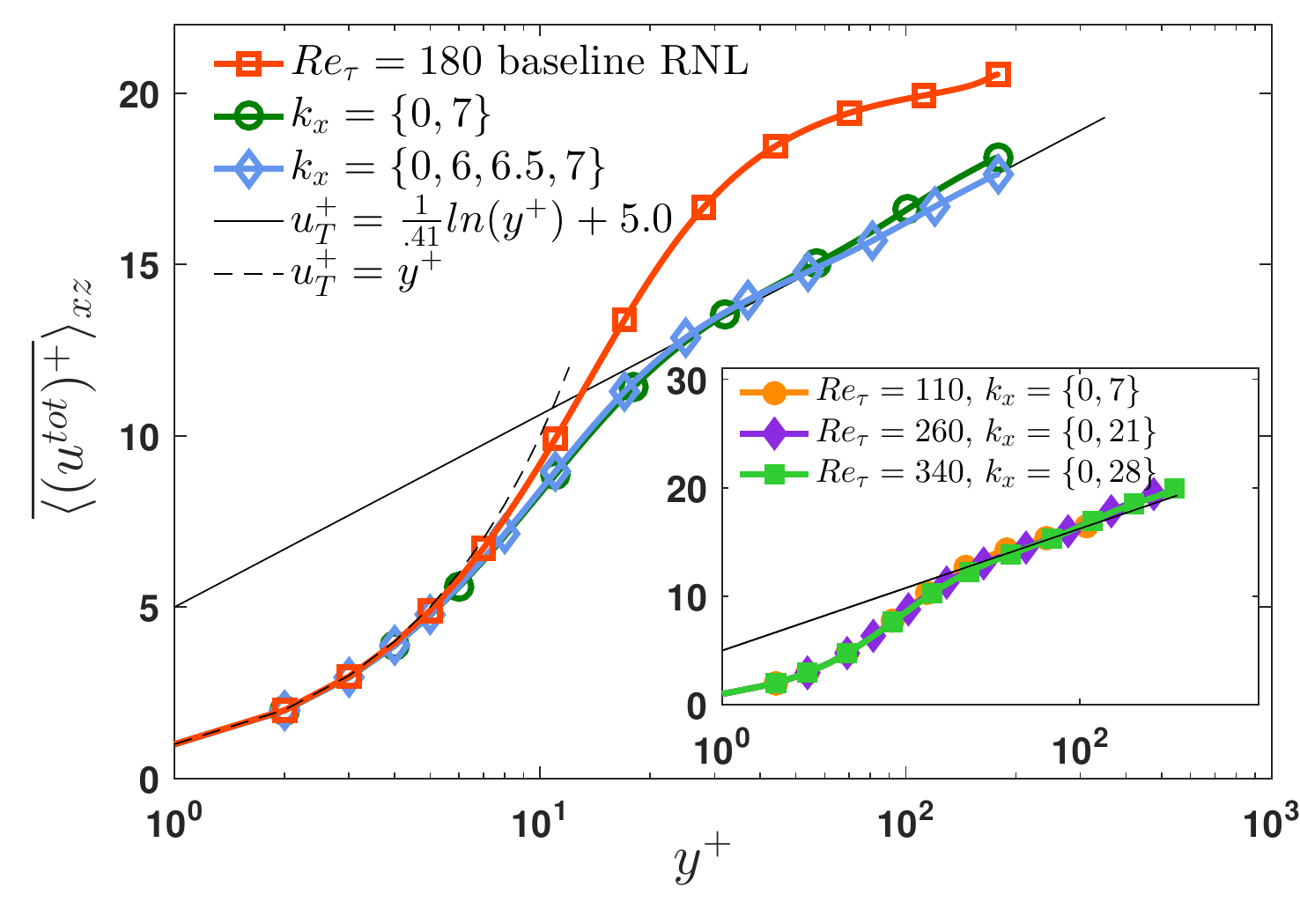}}\hspace{0.025in}
\subfigure[\label{fig:3d}]{
\includegraphics[width = 0.46\textwidth,clip=]{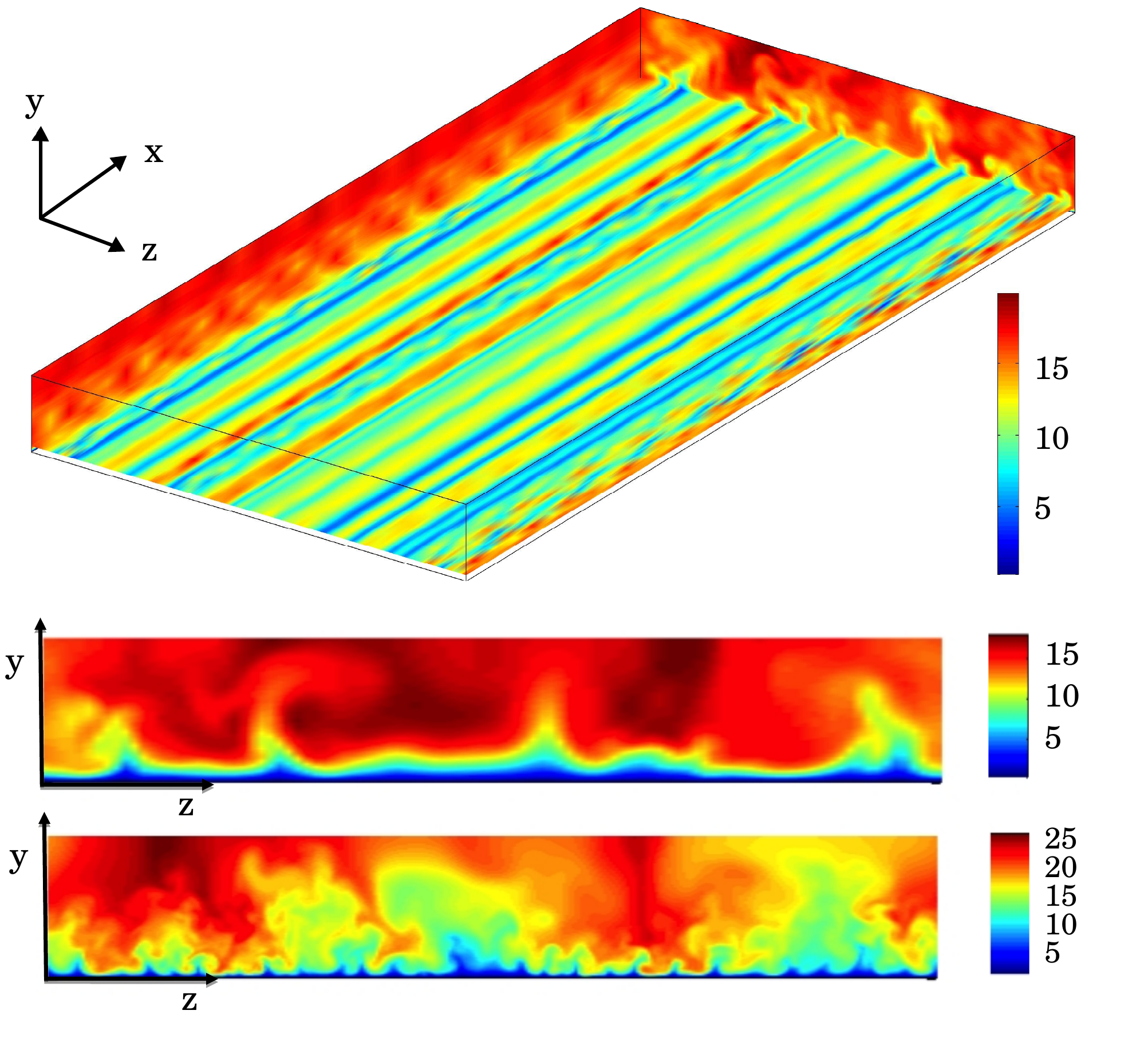} }\vspace{-5pt}
\caption{ \label{fig:half_channelprofiles} (a) Mean velocity profiles for baseline and band-limited RNL simulations  at $Re_\tau=180$ in the forefront,  and band-limited simulations at $Re_\tau=110,\,260$ and $340$ in the inset. (b) (Top) Snapshot of the streamwise velocity at a horizontal plane of $y^{+} = 15$  for a band-limited RNL flow at Re$_{\tau} = 180$ with only $k_x = \{0, 6, 13, 14\}$ . Cross-stream snapshots at Re$_{\tau} = 110$ (center) and Re$_{\tau} = 340$ (bottom) with respective streamwise wave numbers support sets of $k_x = \{0,7\}$ and   $k_x=\{0,28\}$.  This figure is adapted from \cite{Bretheim-etal-2015}, where the wave numbers {\color{black} reported there have been rescaled} so that $k_x=1$ corresponds to $\lambda_x=4\pi \delta$ {\color{black} in order to be consistent with the previous section.}}\vspace{-16pt}
  \end{figure*}

Previous studies of RNL simulations in Pouseuille flow (a full channel) with  $\varepsilon=0$   have demonstrated that the accuracy of the mean velocity profile degrades as the Reynolds number is increased \cite{Constantinou-etal-2014b, Farrell-etal-2016-VLSM}.  This deviation from the DNS mean velocity profile is also seen in simulations of a half channel at $Re_\tau=180$, as shown in Figure \ref{fig:mvps}.  However, the previously observed ability to modify the flow properties through band-limiting the perturbation field can be exploited to improve the accuracy of the RNL predictions. Mean velocity profiles from a series of band-limited RNL simulations at Reynolds numbers ranging from $Re_\tau=180$ to $Re_\tau=340$
in which the improved
accuracy over baseline RNL simulations is clear
are shown in Figure \ref{fig:mvps}.  In particular,  the mean profiles over
this Reynolds number range exhibit a logarithmic region with standard values of $\kappa=0.41$ and $B=5.0$. It should also be noted that many of these band-limited RNL simulations have perturbation fields that are supported by a single streamwise varying wave number, although increasing the support to include a set of three adjacent $k_x\neq 0$ wave numbers
results in slightly improved statistics at $Re_{\tau}=180$.  Similar improvements are seen
in the second-order statistics as reported in \cite{Bretheim-etal-2015}.
The specific wave number to be retained in the model  in order to produce the results
shown here were determined  empirically by comparing the skin friction coefficient  of the band-limited RNL profiles and those obtained from a well validated DNS \cite{Moser99}. {\color{black} That work demonstrated that the wave length producing the best fit over the range of Reynolds numbers shown scales with Reynolds number and asymptotes to a value of approximately $\lambda_x=150$ wall units.  Preliminary work at higher Reynolds numbers has shown that this trend appears to continue to higher Reynolds numbers, although multiple wave numbers (of the same approximate wave length) may be needed.  Developing the theory underlying this behavior is a direction of continuing work.}

\begin{figure}[htbp]
\begin{center}
\includegraphics[width=0.49\linewidth]{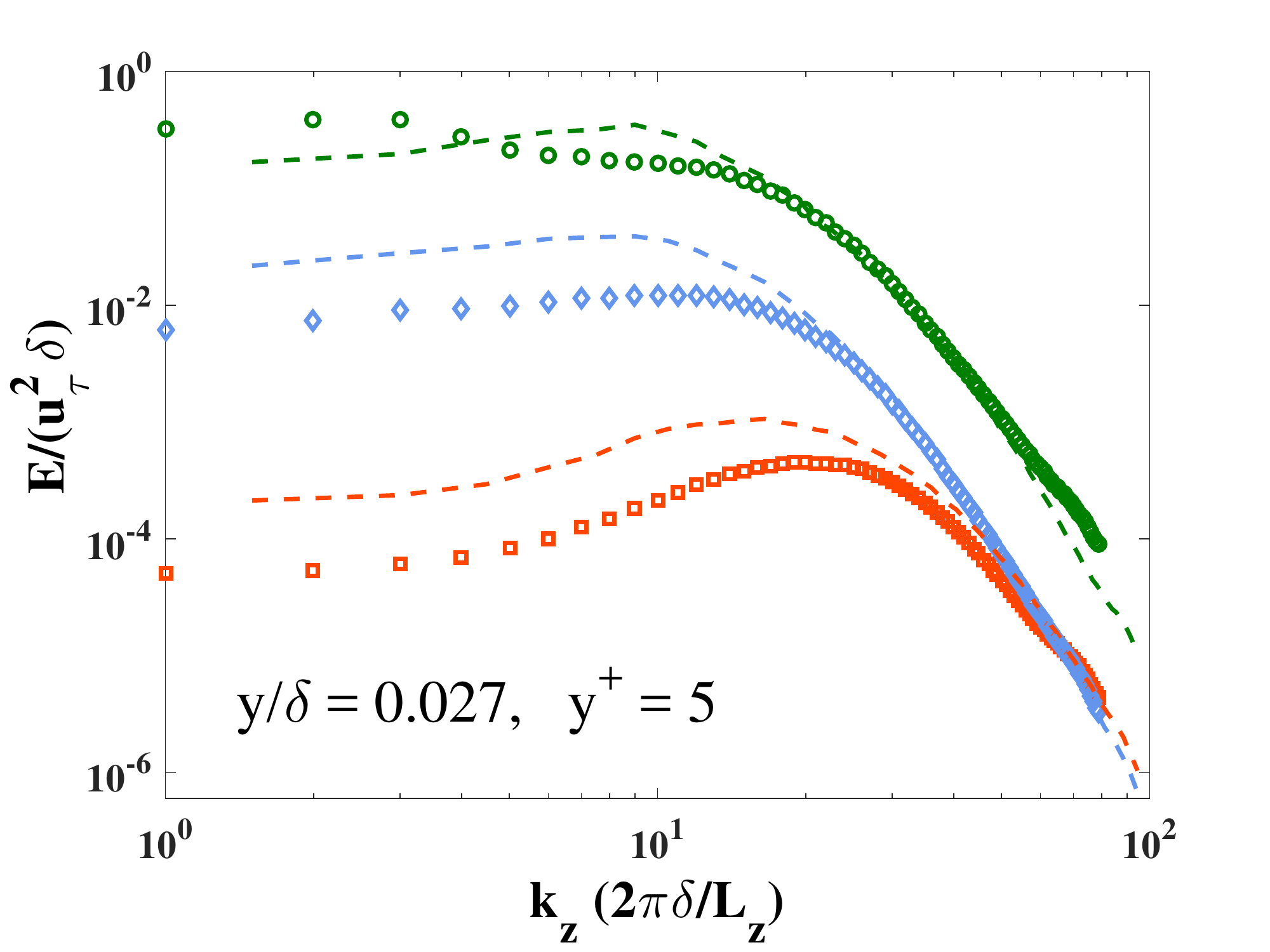} 
\includegraphics[width=0.49\textwidth]{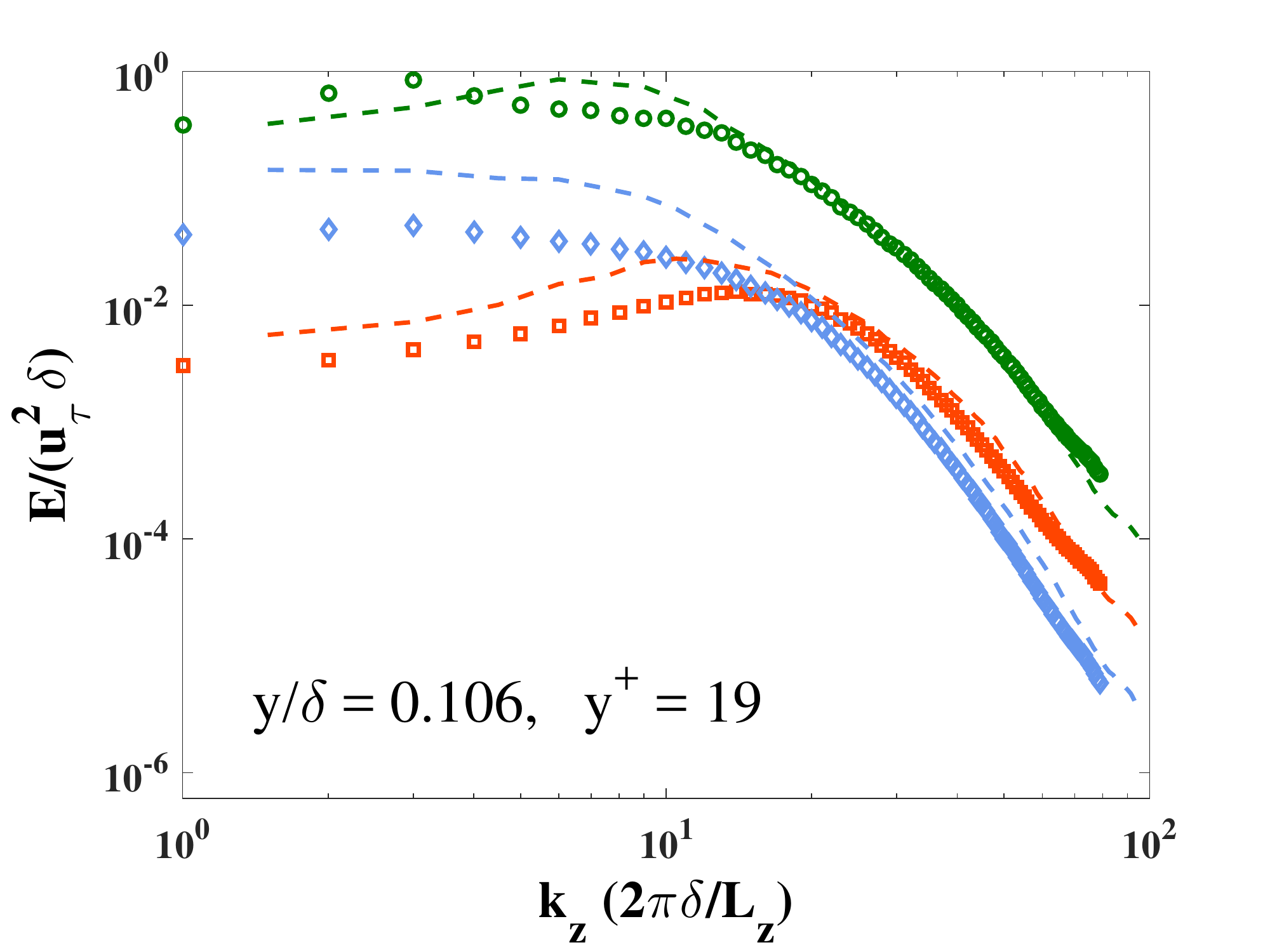}\\ 
(a) \hspace{0.25\linewidth} (b)
\end{center}
\caption{Spanwise energy spectra, {\color{black} $E_{uu}$ (green circles), $E_{vv}$ (red squares) and $E_{ww}$ (blue diamonds),} obtained from the band-limited RNL model at Re$_{\tau}=180$, at two wall-normal locations. The RNL system is constrained to a single perturbation wave number of $k_x=7$. Dashed lines are channel flow DNS data from Moser et al.\cite{Moser99} Symbols are RNL data. This figure is from \cite{Bretheim-etal-2015}.}
\label{fig:spectra}
\end{figure}

Figure \ref{fig:3d} shows snapshots of the streamwise velocity fields for three of the band-limited RNL flows shown in  Figure \ref{fig:mvps}.  The top image shows a horizontal ($x-z$) plane  snapshot of the streamwise velocity, $u^{tot}$, at $y^{+} = 15$ at $Re_{\tau} = 180$ while the middle and bottom images depict cross plane ($y-z$) snapshots of the flow fields at $Re_{\tau} = 110$ and $Re_{\tau} = 340$, respectively. These images demonstrate  realistic vortical structures in the cross-stream,  while the band-limited nature of the streamwise-varying perturbations and the associated restriction to a particular set of streamwise wavelengths is clearly visible in the horizontal plane.  The agreement of the transverse spatial structure of the fluctuations can be quantified through the comparisons of the spanwise spectra with DNS shown in Figure \ref{fig:spectra}.  Here we report results at two distances from the wall for the $Re_\tau=180$ data for the band-limited RNL simulation supported by a perturbation field limited to $k_x = 14$ and a DNS at the same Reynolds number \cite{Moser99}.  Although there are some differences in the magnitudes of the spectra, especially at low wave numbers,  the qualitative agreement is very good considering the simplicity of the RNL model compared to the NS equations.   The benefit of the RNL approach is that these results are obtained at a significantly reduced computational costs.  

\section{Conclusion}

Adopting the perspective of statistical state dynamics (SSD) provides not only new concepts and new methods for studying the dynamics underlying
wall-turbulence but also new reduced order models for simulating wall-turbulence.
The conceptual advance arising from SSD that we have reviewed here is the existence of analytical structures
underlying turbulence  dynamics that lack expression in the dynamics of realizations.
The example we provided is that of the analytical unstable eigenmode and associated bifurcation
structure associated with insatiability of the roll/streak structure in boundary layers subject to which has no analytical expression in the dynamics of realizations.  The modeling advance that we reviewed
is the naturally occurring reduction in order of  RNL turbulence that allows construction of low dimensional models
for simulating turbulence. These models are obtained through a dynamical restriction of the NS equations that forms a SSD or an approximation based on a finite number of realizations of the perturbation field all having a common mean flow, the restricted nonlinear RNL model. A RNL system with an infinite number of realizations, referred to as S3T, provides the conceptual advance, while the RNL approximation provides an efficient computational tool.  The computational simplicity and the ability to band-limit the streamwise wave number support to improve the accuracy means that RNL simulations promise to provide a computationally tractable tool for  probing the dynamics of high Reynolds number flows.  The SSD perspective provides a  set of tools that can provide new insights into wall-turbulence.

\enlargethispage{20pt}

%
%
\aucontribute{The author order is alphabetical. Sections \ref{sec:intro}-\ref{sec:S3T} were primarily written by BFF and PJI with input by DFG.  Sections \ref{sec:RNL_selfsustain} and \ref{sec:RNL_halfchannel} were written by DFG with input by BFF and PJI.  }
%
%
\funding{Partial support from the National Science Foundation under AGS-1246929 (BFF) and a JHU Catalyst Award (DFG) is gratefully acknowledged.}
\ack{The authors would like to thank Charles Meneveau, Navid Constantinou and Vaughan Thomas for a number of helpful discussions and insightful comments on the manuscript.}
%

\end{document}